\documentclass[aps,prl,twocolumn,superscriptaddress,nobibnotes]{revtex4-2}
\pdfoutput=1
\usepackage{graphicx}
\usepackage{amsmath}
\usepackage{amssymb}
\usepackage{siunitx}
\usepackage[version=4]{mhchem}
\usepackage{lmodern}
\usepackage{gensymb}
\usepackage{xcolor}
\usepackage{color}
\usepackage{cleveref}
\usepackage{ragged2e}

\makeatletter
\def\section{%
  \@startsection{section}{2}{\z@}%
    {.6cm \@plus1ex \@minus .2ex}
    {.375cm}
    {\normalfont\small\bfseries\centering}}
\makeatother

\makeatletter
\newcommand{\SItocHide}{%
  \let\SItoc@l@section\l@section
  \let\SItoc@l@subsection\l@subsection
  \let\SItoc@l@subsubsection\l@subsubsection
  \let\l@section\@gobbletwo
  \let\l@subsection\@gobbletwo
  \let\l@subsubsection\@gobbletwo}
\newcommand{\SItocShow}{%
  \let\l@section\SItoc@l@section
  \let\l@subsection\SItoc@l@subsection
  \let\l@subsubsection\SItoc@l@subsubsection}
\makeatother

\begin{document}

\title{Dynamical many-body control of high harmonic generation in a Mott insulator}

\author{Mingyao Guo}
\email[Contact author: ]{mguo@caltech.edu}
\affiliation{Institute for Quantum Information and Matter, California Institute of Technology, Pasadena, California 91125, USA}
\affiliation{Department of Physics, California Institute of Technology, Pasadena, California 91125, USA}

\author{Ryo Noguchi}
\affiliation{Institute for Quantum Information and Matter, California Institute of Technology, Pasadena, California 91125, USA}
\affiliation{Department of Physics, California Institute of Technology, Pasadena, California 91125, USA}

\author{Mohammad Muntasir Hassan}
\affiliation{Institute for Quantum Information and Matter, California Institute of Technology, Pasadena, California 91125, USA}
\affiliation{Department of Physics, California Institute of Technology, Pasadena, California 91125, USA}

\author{Steven J. Gomez Alvarado}
\affiliation{Materials Department, University of California, Santa Barbara, California 93106, USA}

\author{Reina M. Gomez}
\affiliation{Materials Department, University of California, Santa Barbara, California 93106, USA}

\author{Yao-Jui Chan}
\affiliation{Institute of Physics, Academia Sinica, Taipei 11529, Taiwan}

\author{Yu-Chieh Wen}
\affiliation{Institute of Physics, Academia Sinica, Taipei 11529, Taiwan}

\author{Stephen D. Wilson}
\affiliation{Materials Department, University of California, Santa Barbara, California 93106, USA}

\author{Yao Wang}
\email[Contact author: ]{yao.wang@emory.edu}
\affiliation{Department of Chemistry, Emory University, Atlanta, Georgia 30322, USA}

\author{David Hsieh}
\email[Contact author: ]{dhsieh@caltech.edu}
\affiliation{Institute for Quantum Information and Matter, California Institute of Technology, Pasadena, California 91125, USA}
\affiliation{Department of Physics, California Institute of Technology, Pasadena, California 91125, USA}

\begin{abstract}
High harmonic generation (HHG) has emerged as a powerful probe of ultrafast electron dynamics in solids. However, in strongly correlated materials, where electronic excitations arise from many-body interactions rather than single-particle band dispersion, the field remains in its infancy. Here, we report HHG up to the 23$^{\textrm{rd}}$ order in the prototypical Mott insulator La$_2$CuO$_4$. We observe a robust 4~eV cutoff that is independent of driving field strength, revealing an intrinsic energy scale defined by the Hubbard bands. Using time-resolved high harmonic spectroscopy, we disentangle the distinct contributions of carrier creation, intraband acceleration, and interband recombination to the nonequilibrium radiation spectrum. We demonstrate that photo-doping triggers a fundamental reorganization of the many-body manifolds above a critical excitation density, a process that renormalizes the band curvature and suppresses nonlinear intraband currents, thereby directly imprinting the signature of many-body band reconstruction onto the harmonic emission. These results establish HHG as a sensitive probe of subtle correlation-driven electronic reconstructions and highlight that many-body interactions enable extreme nonlinear optical responses that are highly tunable in nonequilibrium states.
\end{abstract}
\maketitle
\addtocontents{toc}{\protect\SItocHide}

\section{Introduction}
High harmonic generation (HHG) is a strong-field phenomenon initially discovered in atomic targets that has laid the foundations of attoscience \cite{krausz2009attosecond}. It has since been observed in various crystalline solids \cite{ghimire2019high,goulielmakis2022high}, emerging as an ultrafast technique to study materials properties ranging from band structure and quantum geometry to electron dynamics and phase transitions \cite{zong2023emerging}. Recently, HHG has been observed in several strongly correlated quantum materials \cite{bionta2021tracking,uchida2022high,alcala2022high,nakano2024dominant,ikeda2025high}; however, the role of many-body correlations in shaping the harmonic emission remains largely unexplored. While these systems possess intrinsic dynamic interactions that make their electronic states highly susceptible to external perturbations, the potential to harness or probe this sensitivity through strong-field processes has not yet been realized. Here, we show that extending HHG spectroscopy to this regime opens a new frontier for both the all-optical manipulation of harmonic generation and the direct interrogation of ultrafast many-body reconstructions — phenomena with no analog in weakly correlated band insulators \cite{zhang2014dynamics,de2021colloquium}.

Undoped cuprate Mott insulators are paradigmatic strongly correlated electron systems. The on-site Coulomb repulsion among their Cu 3$d_{x^2-y^2}$ orbitals splits the electronic states near the Fermi level into lower (LHB) and upper Hubbard bands (UHB), with the former lying below O 2$p$ bands (left inset, Fig.~\ref{fig:m1}A). Consequently, the lowest-energy excitation corresponds to a charge transfer (CT) between occupied oxygen and unoccupied copper states, giving rise to a sizable CT gap in the visible regime \cite{uchida1991optical,falck1992charge}. This large gap permits the application of atomically strong infrared (IR) lightwaves (peak intensity $\sim$1~TW/cm$^2$) across a wide frequency range, rendering the system well-suited for strong-field excitation. Furthermore, due to the formation of Zhang-Rice singlets (ZRS) upon doping, its electronic properties can be effectively captured by a single-band Hubbard model, allowing for a tractable theoretical description \cite{zhang1988effective}. Together, these attributes make parent cuprates a model platform to investigate universal features of HHG in Mott insulators and to demonstrate how light-induced changes to the many-body interacting state can be used to control harmonic dynamics.

\section{HHG from a cuprate Mott insulator}
\begin{figure*}[!htb]
	\centering
	\includegraphics[width=\textwidth]{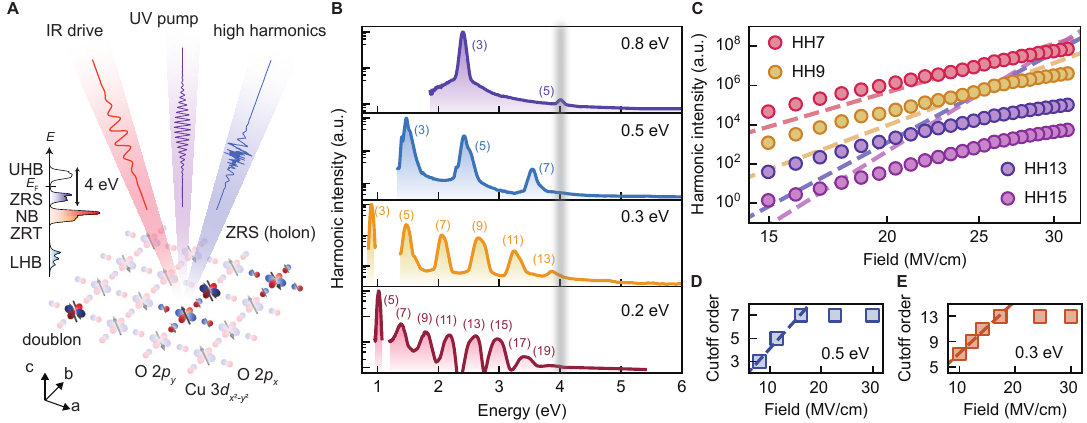}
	\caption{\textbf{HHG from bulk La$_2$CuO$_4$.}
		(\textbf{A}) Schematic of experimental geometry and microscopic processes of HHG in a cuprate Mott insulator. The left inset shows a sketch of density of states at different energy hierarchies, and equilibrium occupied states are filled with color. NB, non-bonding oxygen band; ZRT, Zhang-Rice triplet. (\textbf{B}) HHG intensity spectra from a La$_2$CuO$_4$ single crystal at various driving photon energies on a logarithmic scale. The $n$th harmonic orders are labeled as ($n$). The grey shade marks the global cutoff energy around 4~eV. (\textbf{C}) Dependence of the strength of individual harmonics on the peak field of a 0.2~eV driving laser on a log-log scale. Different orders are rescaled for clarity. Dashed lines are fits of $I_n$ to $E^{2n}$, where $n$ is the harmonic order. (\textbf{D}) Relation between the high-energy cutoff and the peak field of a 0.5~eV driving laser. The dashed line shows linear regression of the cutoff order at low field strengths. (\textbf{E}) Same as (\textbf{D}) but with a 0.3~eV driving laser.}
	\label{fig:m1}
\end{figure*}

Intense near- to mid-IR laser pulses ($\sim$100~fs) were used to drive HHG from bulk single crystals of La$_2$CuO$_4$, a canonical square-lattice cuprate Mott insulator. Pulses were linearly polarized along the crystalline $a$-axis and focused onto the (001) surface, reaching a peak intensity of up to 1.28~TW/cm$^2$ (peak field $E = 31$ MV/cm in vacuum) without imposing physical damage (Appendix~\ref{sec:trhhg}). Notably, we collected HHG responses in reflection geometry \cite{vampa2018observation} (Fig.~\ref{fig:m1}A), avoiding nonlinear propagation effects that may distort spectral properties in transmission studies \cite{xia2018nonlinear}. The resulting spectra, shown in Fig.~\ref{fig:m1}B for various driving wavelengths at a fixed peak applied intensity of 1.2~TW/cm$^2$, consist exclusively of odd-order harmonics, due to the inversion symmetry of the La$_2$CuO$_4$ lattice \cite{PhysRevB.104.014304}. The yield of these harmonics deviates sharply from the perturbative power law ($I_n\propto E^{2n}$, dashed lines in Figs.~\ref{fig:m1}C and~\ref{sfig:field dependence} \cite{SI}), establishing the fundamentally non-perturbative nature of the emission.

By systematically varying the driving frequency, we identified an unusual phenomenon: the emergence of a sharp, global cutoff at around 4~eV, regardless of the driving frequency (grey shade, Fig.~\ref{fig:m1}B; Figs.~\ref{sfig:wavelength dependence} and~\ref{sfig:wavelength dependence dense} \cite{SI}). For peak fields below approximately 18~MV/cm, the harmonic cutoff order scales linearly with field strength (Fig.~\ref{fig:m1}D, E), as expected for typical solids \cite{ghimire2011observation,luu2015extreme}. However, as the photon energy approaches this 4~eV threshold, further increases in field strength fail to extend the emission to higher energies. While a constant cutoff can appear in liquid-phase HHG due to a limited electron excursion length \cite{mondal2023high,mondal2026multi}, we ruled this out as the origin here (Appendix~\ref{sec:excursion}; Supplemental Material Sec.~\ref{sec:finite excursion length} \cite{SI}), suggesting instead a mechanism intrinsic to the material's correlated electronic structure. 

HHG in undoped cuprates originates from the nonlinear charge dynamics within the many-body manifolds \cite{ikeda2025high}. While strong correlations preclude single-particle descriptions, the three-step model for HHG still provides useful insights when generalized to the Mott insulating state \cite{murakami2018high,murakami2021high}. Here, tunneling across the CT gap creates doublons (doubly occupied Cu sites) and holons, the latter manifesting as ZRS formed by the entanglement of a Cu hole and an adjacent O photo-hole \cite{zhang1988effective} (Fig.~\ref{fig:m1}A). The subsequent coherent hopping and recollision of these quasiparticles generate high harmonics, analogous to intraband Bloch oscillations and interband recombination in semiconductors respectively \cite{ghimire2019high,goulielmakis2022high}. Crucially, the maximum energy for doublon-holon recombination is bounded by the separation between the ZRS bottom and the UHB top — a scale established at $\sim$4~eV in cuprates \cite{chen1991electronic,ye2013visualizing,damascelli2003angle}. The alignment of this energy scale with our observed cutoff suggests that the harmonic generation process is fundamentally limited by recollisions within these many-body states across the Mott gap. The absence of secondary plateaus \cite{ndabashimiye2016solid,uzan2020attosecond,mondal2026multi} or multi-band effects \cite{schubert2014sub,yue2022signatures} in our measured spectra further indicates that weakly correlated non-bonding oxygen bands play a minimal role in the emission.

\section{Time-resolved HHG spectroscopy}
\begin{figure*}[!htb]
	\centering
	\includegraphics[width=\textwidth]{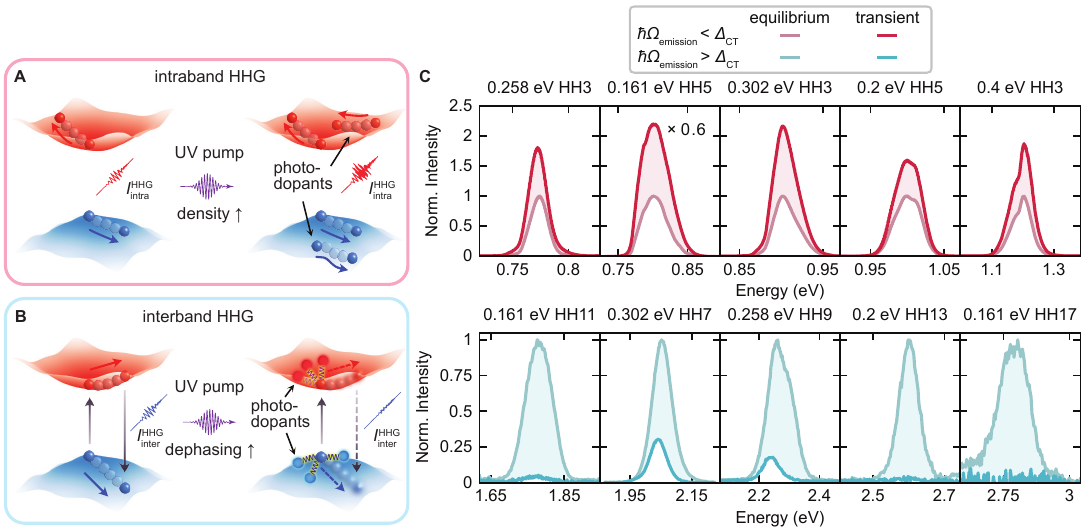}
	\caption{\textbf{Time-resolved HHG spectroscopy of La$_2$CuO$_4$.}
		(\textbf{A}), (\textbf{B}) Schematic of how intra- (\textbf{A}) and inter-band (\textbf{B}) HHG processes respond to photo-excited carriers in momentum space. The left and right panels show the dynamics before and after photo-doping, respectively. (\textbf{C}) Transient HHG spectra following a 1 mJ/cm$^2$ 3.1~eV pump normalized by their equilibrium maximal intensity at time delay $t= 300$~fs. The upper and lower panels display the responses of below- and above-gap high harmonics, respectively.}
	\label{fig:m2}
\end{figure*}

We next seek to identify the microscopic origin of the observed harmonics and realize active control of their emission. Previous studies on Mott insulators \cite{graanas2022ultrafast,uchida2022high,ikeda2025high} relied on a single IR pulse to both generate charge carriers and drive their subsequent dynamics, inherently conflating the contributions of intraband and interband processes (Fig.~\ref{fig:m2}A,B). Time-resolved HHG spectroscopy provides a powerful route to overcome this limitation \cite{van2024toward,heide2024ultrafast}. By decoupling a certain amount of carrier excitation from harmonic generation — specifically through photo-doping with an above-gap pump pulse — we can selectively probe individual pathways. Photocarrier injection is expected to amplify the intraband current by increasing the density of doublons and holons available to be driven by the IR field, potentially augmenting the HHG yield (Fig.~\ref{fig:m2}A). Conversely, these injected carriers are mutually phase-incoherent so they suppress interband recombination through excitation-induced dephasing, a density-dependent scattering process (Fig.~\ref{fig:m2}B). By evaluating the ultrafast response of harmonics following photo-doping, we can thus distinguish these two competing contributions. 

We generated a nonequilibrium population of doublons and holons using a 3.1~eV pump pulse (duration 150~fs, see Appendix~\ref{sec:trhhg}), far exceeding the CT gap of La$_2$CuO$_4$ \cite{uchida1991optical,falck1992charge}. Figure~\ref{fig:m2}C displays HHG spectra across a range of driving wavelengths before and immediately after pump excitation at a fluence of 1~mJ/cm$^2$ (see Fig.~\ref{sfig:hhg dynamics} for full dynamics \cite{SI}). Strikingly different responses are observed in different emission regimes. While low-order harmonics below the CT gap generally exhibit increased yields, high-order harmonics above the gap are significantly suppressed. The suppression becomes more pronounced at higher harmonic orders (Fig.~\ref{sfig:harmonic order dependence} \cite{SI}), which are expected to be more susceptible to dephasing. 

Because photo-doping enhances intraband currents but destroys interband coherence, these divergent responses provide clear evidence for the respective intraband and interband origins of below- and above-gap harmonics. This assignment is consistent with our observation of a constant cutoff energy and aligns with theoretical predictions from dynamical mean-field theory \cite{murakami2018high,ikeda2025high} and density matrix renormalization group studies \cite{murakami2021high}. Furthermore, we observe a substantial enhancement of below-gap HHG — exceeding a factor of four when driven at 0.161~eV at high fluences (Figs.~\ref{fig:m2}C and~\ref{sfig:hh5 enhancement} \cite{SI}) — representing the highest amplification reported to date in solid-state systems \cite{nagai2023effect,xu2023defect,suthar2024momentum}. This measured four-fold increase likely represents a lower bound; accounting for the penetration depth and coherence length mismatch between the ultraviolet (UV) pump and IR harmonic pulses, the intrinsic enhancement is expected to be more than one order of magnitude (Supplemental Material Sec.~\ref{sec:giant intraband HHG} \cite{SI}), underscoring the remarkable potential for optical control of HHG emission pathways and yield in these systems.

\begin{figure*}[htb]
	\centering
	\includegraphics[width=\textwidth]{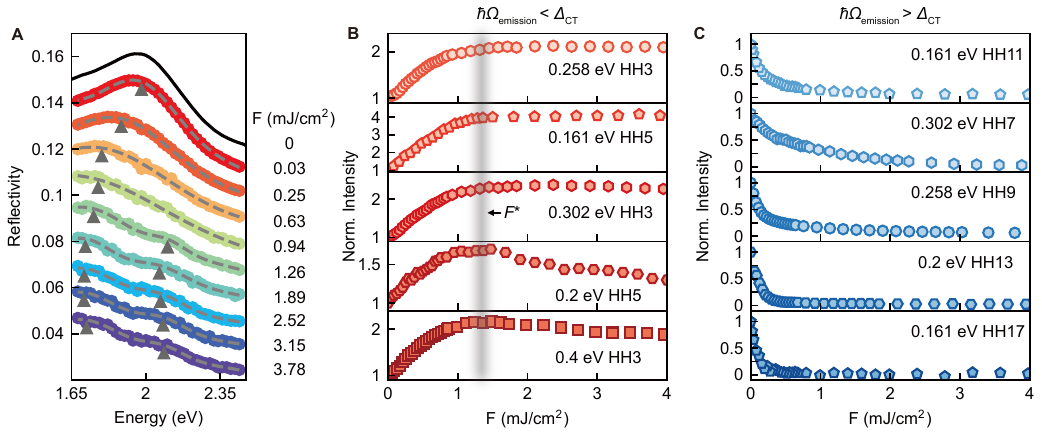}
	\caption{\textbf{Light-induced many-body band renormalization reshapes HHG from La$_2$CuO$_4$.}
		(\textbf{A}) Broadband transient reflectivity spectra at time delay $t=300$~fs for various 400~nm pump fluences. The topmost black curve displays the equilibrium spectrum. Local peak positions extracted from fits (dashed lines) are labeled by triangles (see Supplemental Material Sec.~\ref{fitting R} for details of the fitting procedure \cite{SI}). Curves are vertically offset for clarity. (\textbf{B}) Pump fluence dependence of the integrated transient below-gap HHG intensity at $t=300$~fs. The grey shade indicates $F^*$ determined from broadband reflectivity measurements. (\textbf{C}) Same as (\textbf{B}) but for above-gap HHG.}
	\label{fig:m3}
\end{figure*}

\section{Many-body control of HHG out of equilibrium}
Having utilized photo-doping to disentangle the microscopic channels of HHG, we next explore how light-induced reorganization of the many-body landscape — a phenomenon inherently rooted in strong electron correlations — fundamentally alters the emission process at higher photo-doping levels. Because intraband harmonics are governed by the dispersion of the correlated manifolds, they should be exquisitely sensitive to any renormalization of many-body excitations. To investigate this, we utilized broadband transient reflectivity to probe the electronic state under intense excitation (Appendix~\ref{sec:trr}). In equilibrium, the reflectance spectrum is dominated by a single peak corresponding to the CT transition \cite{uchida1991optical,falck1992charge}. Immediately following 3.1~eV pump excitation, we observe the expected suppression and slight redshift of this peak (Fig.~\ref{fig:m3}A), consistent with standard optical bleaching and dynamical screening effects \cite{okamoto2011photoinduced,novelli2014witnessing,baldini2020electron,huang2025ultrafast}. 

However, as the pump fluence crosses a critical threshold of $F^*\approx1.25$ mJ/cm$^2$, the CT resonance bifurcates into two distinct peaks (Fig.~\ref{fig:m3}A). Neither heating \cite{falck1992charge} nor dielectric screening \cite{baykusheva2022ultrafast} can account for this behavior, as both mechanisms typically reduce the gap size without generating new resonances. Furthermore, the effect occurs well below the threshold for a structural phase transition \cite{gedik2007nonequilibrium}. While a similar splitting occurs upon chemical doping in La$_{2-x}$Sr$_x$CuO$_4$ \cite{uchida1991optical,kim2021optical}, where introducing holes redistributes spectral weight between emergent quasiparticle bands and the remnant Hubbard bands, its appearance here occurs at a different energy scale, suggesting that nonequilibrium photo-doping is fundamentally distinct from its equilibrium counterpart. We therefore attribute this behavior to a photo-doping-induced reorganization of the many-body manifolds. The critical fluence we observe lies well above the excitation densities at which transient metallicity has previously been reported in photo-doped La$_2$CuO$_4$ \cite{okamoto2011photoinduced,novelli2014witnessing,baldini2020electron}, indicating that the reorganization only occurs once a substantially high density of itinerant photocarriers is achieved.

To further corroborate this nonequilibrium reconstruction and examine its influence on HHG, we monitored the harmonic emission normalized to its unpumped intensity (Fig.~\ref{fig:m3}B,C). In the weak excitation regime (below 1~mJ/cm$^2$), the below-gap harmonics — associated with the intraband motion of doublons and holons — exhibit a linear response to the injected photocarrier density. However, as the fluence approaches $F^*$, the HHG enhancement at all wavelengths begins to level off or even decline (Fig.~\ref{fig:m3}B), reflecting the dramatic electronic changes observed in our reflectivity data. 

\begin{figure*}[!htb]
	\centering
	\includegraphics[width=\textwidth]{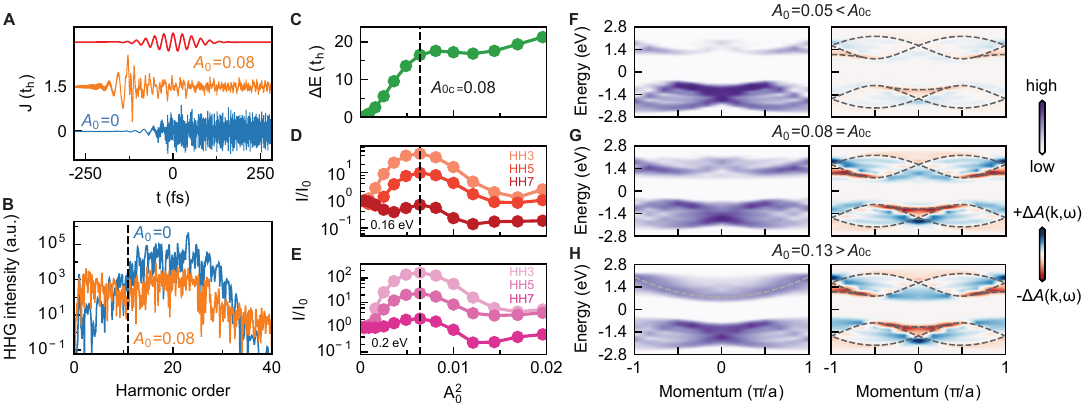}
	\caption{\textbf{Hubbard model numerics.}
		(\textbf{A}) Current in the absence ($A_0=0$, blue line) and presence ($A_0=0.08$, orange line) of the pump field under the $\hbar\omega=0.16$~eV mid-IR driving pulse (red line). Different curves are vertically offset for clarity. (\textbf{B}) HHG spectra of the mid-IR field in the unpumped and pumped conditions same as in (\textbf{A}). The dashed line denotes the Mott gap energy. (\textbf{C}) Pump fluence dependence of pump-injected energies at the end of the pump pulse. (\textbf{D}) Pump fluence dependence of below-gap HHG with $\hbar\omega=0.16$~eV. (\textbf{E}) Same as (\textbf{D}) but with $\hbar\omega=0.2$~eV. (\textbf{F})--(\textbf{H}) Single-particle spectral functions $\mathcal{A}(k,\omega)$ below (\textbf{F}), near (\textbf{G}), and above (\textbf{H}) the threshold field at the end of the pump pulse. The left column shows the removal function and a logarithmic color scale is adopted to make excitations in the UHB discernible. The grey dashed line in (\textbf{H}) shows the fit of renormalized UHB dispersion to a parabolic function. The right column shows the difference between transient and static total spectral function and the color scale is linear. The grey dashed lines in each right panel show the spinon and holon dispersion in equilibrium.}
	\label{fig:m4}
\end{figure*}

As discussed in Supplemental Material Secs. \ref{sec:absorption saturation} and \ref{sec:eid} \cite{SI}, the absorption saturation and excitation-induced dephasing play a minor role in this regime. Instead, the saturation of the HHG yield aligns with a fundamental renormalization of the correlated electronic structure: as the photo-doping level rises beyond a threshold, spectral weight from the equilibrium Hubbard bands reorganizes into a transient dispersion that suppresses nonlinear intraband motion. This strong correspondence between transient reflectivity and HHG yield provides a clear demonstration of many-body control over HHG in La$_2$CuO$_4$. In turn, these results establish HHG spectroscopy as a sensitive probe of correlation-driven band renormalization on ultrafast timescales.

In contrast, the above-gap harmonics — dominated by interband recombination between the UHB and ZRS — remain relatively insensitive to this many-body renormalization, evolving smoothly across $F^*$ (Figs.~\ref{fig:m3}C and \ref{sfig:harmonic order dependence} \cite{SI}). This is consistent with our reflectivity data showing that the CT gap remains relatively rigid in this regime. Their monotonic suppression is well-captured by a phenomenological model incorporating the excitation-induced dephasing process (Supplemental Material Sec.~\ref{sec:saturation model} \cite{SI}). Ultimately, the divergent sensitivity of below- and above-gap harmonics highlights how the specific many-body nature of the electronic states dictates the nonlinear optical response.

\section{Hubbard model simulation}
To understand the light-induced many-body reconstruction of the Mott insulating state on a microscopic level, we modeled the system using a half-filled single-band Hubbard Hamiltonian. We numerically solved the time-dependent Schr{\"o}dinger equation via exact diagonalization on a 12-site chain (Appendix~\ref{sec:tded}), choosing parameters $U=8t_h$ and $t_h=0.35$~eV to mirror the on-site Coulomb repulsion and inter-site hopping energy scales of cuprates \cite{pavarini2001band,yoshida2006systematic}. This model captures the essential physics of doublon–holon dynamics within the UHB and ZRS manifolds, allowing us to resolve the nonequilibrium evolution of the electronic state upon photo-doping.

In our simulation, we matched experimental conditions with an IR drive of $\hbar\omega=0.16$~eV and a UV pump of $\hbar\Omega=3.1$~eV. The microscopic current driven by the IR pulse alone exhibits prominent nonlinear responses that give rise to HHG (blue curves, Fig.~\ref{fig:m4}A,B). Upon photo-excitation across the Mott gap, the time-domain current changes dramatically: low-frequency components are enhanced while high-frequency oscillations are suppressed (orange curves, Fig.~\ref{fig:m4}A,B). This crossover energy aligns closely with the Mott gap (Fig.~\ref{fig:m4}B), and our calculated time-frequency spectrogram (Fig.~\ref{sfig:tr hhg spectrum} \cite{SI}) and carrier number dynamics (Fig.~\ref{sfig:hubbard carrier dynamics} \cite{SI}) confirm that these below- and above-gap HHG channels correspond to doublon–holon hopping and recombination, respectively, supporting our experimental assignments. 

As the pump amplitude $A_0$ approaches a critical strength $A_{0c}=0.08$, the normalized intensity of below-gap harmonics starts to exhibit a pronounced nonlinear and even non-monotonic behavior (Fig.~\ref{fig:m4}D). Notably, this threshold $A_{0c}$ coincides with a sharp kink in the pump-injected energy (Fig.~\ref{fig:m4}C), signaling a nonthermal departure from the linear response regime. To corroborate this threshold, we simulated the HHG responses of another driving pulse at $\hbar\omega=0.2$~eV, which shows the same behavior across $A_{0c}$ (Fig.~\ref{fig:m4}E), consistent with the experimentally observed universality of $F^*$ across various IR drive wavelengths (Fig.~\ref{fig:m3}B). While the exact functional form of the fluence dependence in our one-dimensional model differs slightly from the two-dimensional lattice of La$_2$CuO$_4$ due to differing band dispersions and dimensions (Supplemental Material Sec.~\ref{sec:hubbard simulations} \cite{SI}), the qualitative trend remains consistent. 

Building on this correspondence, we analyzed the electronic origin of this threshold behavior by computing the time-dependent spectral function $\mathcal{A}(k,\omega;t)$ at the conclusion of the pump pulse, which directly reveals the structure of single-particle excitations. In equilibrium, $\mathcal{A}(k,\omega)$ is not only gapped into LHB and UHB, but also exhibits incoherent features within each of them (Fig.~\ref{sfig:hubbard spectral function}A \cite{SI}). These incoherences reflect the strongly correlated electrons which lose the quasiparticle structure at high binding energies \cite{kim2006distinct,kohno2010spectral,graf2007universal,wang2020emergence}. At low fluence ($A_0<A_{0c}$), the UHB is sparsely populated with redistributed spectral weight concentrated at energies in resonance with the pump (Fig.~\ref{fig:m4}F, left), without shifting the poles in the density of states. To isolate changes in dispersion, we examined the difference spectrum $\Delta\mathcal{A}(k,\omega)$ between the excited and ground states (Fig.~\ref{fig:m4}F, right). In this regime, $\Delta\mathcal{A}(k,\omega)$ is negligible outside the unperturbed Hubbard bands, indicating that photo-excitation primarily redistributes spectral weight without appreciable band renormalization. 

Approaching the critical field strength ($A_{0c}=0.08$), finite spectral weight begins to accumulate near the $\Gamma$ point ($k=0$) above the Fermi level (Fig.~\ref{fig:m4}G, left) and lies outside the equilibrium holon dispersion (Fig.~\ref{fig:m4}G, right), signaling correlation-driven modifications beyond simple state filling. At high fluence ($A_0>A_{0c}$), the occupied portion of the UHB close to the Fermi level reorganizes into a nearly parabolic, free-electron-like shape (Fig.~\ref{fig:m4}H, left). This transformation can be explained through a simplified doublon-holon picture. Strong short-range antiferromagnetic (AFM) correlations generate an effective confining interaction between photo-excited doublons and holons through the exchange-energy cost associated with their separation. As the pump fluence increases, the pair density grows approximately linearly, resulting in a corresponding linear increase in the total excitation energy (Fig.~\ref{fig:m4}C). However, once the pair density exceeds a critical threshold at $A_{0c}$, the system enters a regime with strongly suppressed AFM correlations and increasingly free propagation of doublons and holons, which develop a nearly parabolic dispersion (Supplemental Material Sec.~\ref{sec:generalizability} \cite{SI}). The pronounced increase of $\Delta\mathcal{A}(k,\omega)$ within the UHB at momenta between $-\pi/2a$ and $\pi/2a$ around 0.7~eV (Fig.~\ref{fig:m4}H, right), together with analyses of density of states (Fig.~\ref{sfig:hubbard dos} \cite{SI}), further substantiates this dynamical band restructuring. Such a reconstruction is detrimental to the nonlinear intraband oscillations responsible for below-gap HHG, providing a microscopic explanation for the observed behaviors. Meanwhile, the Hubbard manifolds persist without a Mott gap collapse owing to the persistence of double-occupancy penalty, concurring with measured reflectivity spectra and corroborating the insensitivity of above-gap HHG to the photo-induced band renormalization (Fig.~\ref{sfig:hubbard noneq drive field dep} \cite{SI}). These results confirm the many-body origin of tunable HHG from La$_2$CuO$_4$.

\section{Discussion and outlook}
Together, our experiments and simulations establish a coherent framework for HHG dynamics in La$_2$CuO$_4$. In this picture, intraband emission represents a competition between increased photocarrier densities — which enhance the yield — and light-induced many-body renormalization, which makes the bands more parabolic and thus disfavors nonlinear current generation. Conversely, interband harmonics remain insensitive to this electronic reorganization but undergo strong suppression due to the excitation-induced dephasing. The ability to manipulate HHG through light-driven band reconstruction is expected to be broadly applicable across a wide range of strongly correlated quantum materials. 

Furthermore, because high harmonic emission in these systems emerges directly from complex many-body charge dynamics, decoding its time-domain traces offers a unique window into coherent many-body excitations and transient nonequilibrium states \cite{silva2018high,murakami2025photoinduced}. A compelling direction for future research will be to extract quantitative information about the dynamical evolution of the electronic state during the HHG process by time-resolving the radiation field \cite{hohenleutner2015real,garg2016multi,uchida2024high}. Such attosecond clocking of many-body dynamics, paired with precise microscopic modeling, could reveal the fundamental timescales of electron correlation.

\section*{Acknowledgments}
We thank H. Kim, Y. He, A. Srinivasan, D. Reis, R. Andrei, and E. Demler for helpful conversations. D.H. acknowledges support for instrumentation from the David and Lucile Packard Foundation and the Institute for Quantum Information and Matter, an NSF Physics Frontiers Center (PHY-1733907). M.G. acknowledges support from the Eddleman Graduate Fellowship. Y.W. acknowledges support from the U.S. Department of Energy, Office of Science, Basic Energy Sciences, under Early Career Award No.~DE-SC0024524. The simulations presented in this paper used resources of the National Energy Research Scientific Computing Center, a U.S. Department of Energy Office of Science User Facility located at Lawrence Berkeley National Laboratory, operated under Contract No. DE-AC02-05CH11231. S.D.W., S.J.G.A., and R.M.G. gratefully acknowledge support via the UC Santa Barbara NSF Quantum Foundry funded via the Q-AMASE-i program under award DMR-1906325. Y.-J.C. and Y.-C.W. were supported by National Science and Technology Council, Taiwan (Grant No. NSTC 114-2112-M-001-023-) and Academia Sinica (Grant No. AS-iMATE-113-14).

\bigskip
M.G. constructed the setup and performed HHG spectroscopy experiments. M.G. conducted numerical simulations supervised by Y.W. R.N. performed broadband transient reflectivity measurements with the help of M.M.H. S.J.G.A. and R.M.G. grew single crystals of La$_2$CuO$_4$ with guidance from S.D.W. Y.-J.C. characterized the equilibrium optical spectrum with supervision from Y.-C.W. M.G., Y.W., and D.H. wrote the manuscript with contributions from all other authors.

\bigskip
The authors declare no competing interests.

\section*{Data availability}
The data that support the findings of this study are available from the corresponding author upon reasonable request.

\section*{Code availability}
The code used for numerical simulations in this study is available from the corresponding author upon reasonable request.

\appendix
\section{Materials}
\label{sec:material}
We employed a standard traveling solvent floating zone crystal growth recipe, using a flux with a molar composition of 4:1 CuO:La$_2$O$_3$ and a polycrystalline feed rod which contained 2\% excess moles of CuO to compensate for volatility and sustain the flux composition during the growth. Both the polycrystalline feed rod and the flux pellet were prepared via a conventional solid state reaction. Stoichiometric amounts of La$_2$O$_3$ (99.99\%, Alfa Aesar) and CuO (99.9995\% Thermo Scientific) powders were first ground into a homogeneous mixture. Prior to firing, powder was packed into a 5~mm diameter latex balloon and pressed under a hydrostatic pressure of 3275~bar in a cold isostatic press. Material for the feed rod underwent a three-step firing process with intermittent grinding and cold-pressing between each step to improve homogeneity: (1) 800~$^\circ$C for 24~hours in air, (2) 900~$^\circ$C for 24~hours in air, and (3) 1200~$^\circ$C for 12~hours in flowing O$_2$. Material for the flux underwent a similar three-step firing process: (1) 800~$^\circ$C for 24 hours in air, (2) 800~$^\circ$C for 24~hours in air, and (3) 850~$^\circ$C for 12~hours in air. The flux rod was then cut into a disc and placed atop a single crystalline seed of La$_2$CuO$_4$ for floating zone crystal growth in a four-mirror optical floating zone furnace with 300~W halogen lamps (Crystal Systems Corp. Model FZ-T-10000-H-VI-VPO-I-HR-PC). A growth rate of 1~mm/hr was maintained over several days under 1~atm dynamic pressure of flowing O$_2$. 

Typically, the incorporation of a small amount of interstitial oxygen during the crystal growth leads to as-grown crystals which are lightly hole-doped and superconducting. In order to remove the excess oxygen, the following vacuum annealing procedure was performed: heat from 30~$^\circ$C to 910~$^\circ$C over 3~hours, hold for 1~hour, cool to 600~$^\circ$C over 12~hours, cool to 500~$^\circ$C over 200~hours, and finally cool to room temperature over 12~hours \cite{PhysRevB.104.014304}. The resulting crystal was oriented using a Laue X-ray diffractometer and then diced/polished to expose the (001) crystallographic plane (Fig.~\ref{sfig:sample characterization}A \cite{SI}). Magnetic susceptibility measurements were then carried out on a small piece from the annealed crystal using a Quantum Design Magnetic Property Measurement System (MPMS3) under an applied field of 1500~Oe perpendicular to the $c$ axis.
Our magnetization measurements confirm the absence of a low-temperature diamagnetic signal and show a cusp at the AFM ordering temperature $T_\mathrm{N}=325~\mathrm{K}$ expected for the parent Mott insulating state of La$_2$CuO$_4$ \cite{keimer1992neel,PhysRevB.104.014304}, verifying the successful removal of excess interstitial oxygen (Fig.~\ref{sfig:sample characterization}B \cite{SI}).

\section{Equilibrium optical measurements}
\label{sec:reflectivity}
Static reflectance measurements were performed in near normal incidence geometry (incident angle of approximately 10$^\circ$) using a Bruker VERTEX 70v FTIR spectrometer equipped with a liquid helium flow-cryostat. The reflectance spectrum in the photon energy range from 1.115 to 3.10 eV (9000 to 25000 cm$^{-1}$) shown in this work was measured at 295~K under ambient pressure. To obtain the reflectance spectrum, the reflected signal from the sample was first normalized to that of the reference aluminum (Al) mirror. The resulting ratio was then converted to the absolute reflectance through the multiplication with the known reflectance of Al \cite{Rakic1995}, which is displayed as the topmost black curve in Fig.~\ref{fig:m3}A, in agreement with previous optical studies \cite{uchida1991optical}.

Crystal orientation was determined by a high-speed rotational-anisotropy second harmonic generation setup \cite{harter2015high}. An 800-nm laser was used as the probe and the SHG signal was collected in the 'S$_{\rm{in}}$-P$_{\rm{out}}$' geometry (Fig.~\ref{sfig:orientation dependence}A \cite{SI}). Based on the \emph{mmm} point group symmetry of the crystal in its low-temperature orthorhombic phase, we assign the \emph{a} and \emph{b} axes shown in the inset of Fig.~\ref{sfig:orientation dependence}A  \cite{SI}. Because the in-plane anisotropy along the \emph{a} and \emph{b} axes is negligible \cite{PhysRevB.104.014304}, we refer to them as the Cu-O-Cu bond directions without distinguishing them here.

\section{Time-resolved HHG measurements}
\label{sec:trhhg}
The experimental setup (Fig.~\ref{sfig:setup}A \cite{SI}) is based on a femtosecond titanium-sapphire laser amplifier (repetition rate, 1 kHz; pulse duration, 35 fs; center wavelength, 800 nm), which is used to pump a dual-stage optical parametric amplifier (OPA). The output wavelengths of signal/idler pulses can be readily tuned between 1.17 $\mu$m and 2.5 $\mu$m (signal/idler pulse duration, 50/70 fs). Mid-IR pulses are generated by their difference frequency mixing in an AgGaS$_2$ (2.6$\sim$4.5 $\mu$m) or GaSe (4.5$\sim$10 $\mu$m) crystal in a noncollinear geometry (pulse duration, $\sim$100 fs). Each IR pulse wavelength was characterized before the measurements using a home-built grating-based spectrometer (Fig.~\ref{sfig:pulse characterization} \cite{SI}). For mid-IR excitations, a germanium long-pass filter was employed to filter out the residual near-IR pulses. 

High harmonics (HHs) were obtained by focusing the IR beams obliquely (incident angle, 10$^{\circ}$) onto the sample by an off-axis parabolic (OAP) mirror with a 1/$e^2$ beam diameter on the order of 60 $\mu$m depending on the exact excitation wavelength. The highest employed peak driving field of 31 MV/cm is close to the observed long-time damage threshold of La$_2$CuO$_4$. The resulting reflected spectrum was collected by the same OAP mirror and recorded by a home-built grating-based spectrometer with an indium gallium arsenide detector array (1$\sim$1.7 $\mu$m) and a cooled silicon charge-coupled device (0.2$\sim$1.1 $\mu$m). All spectra are corrected for the reflectance of mirrors, the diffraction efficiency of the grating, and the quantum efficiency of the detector.

To generate the UV pump excitation, the fundamental 800 nm pulses were focused onto a 0.2 mm thick type-I $\beta$-barium borate crystal, yielding second harmonic with central wavelength at 400 nm. A hard-coated bandpass filter centered at this wavelength (full-width half maximum, 10 nm) was used to filter out the residual pump pulses. The 400-nm beam was focused onto the sample at normal incidence through a hole in the OAP mirror by a lens with a 1/$e^2$ beam diameter of 150 $\mu$m. A shortpass dichroic beamsplitter (edge wavelength, 414 nm) was employed to filter out the scattered pump photons.

The IR pulses were linearly polarized along the $a$ axis of the $c$-axis oriented bulk La$_2$CuO$_4$ crystal (Supplemental Material Sec.~\ref{sec:orientation} \cite{SI}). The UV pump pulse was linearly polarized along the $b$ axis to minimize nonlinear frequency-mixing signals with the drive field during their temporal overlap. All the HHG measurements were carried out at 295~K.

\section{Transient reflectivity spectroscopy measurements}
\label{sec:trr}
The broadband visible pulses for transient reflectivity spectroscopy (Fig.~\ref{sfig:setup}B \cite{SI}) were produced by supercontinuum generation in a 4-mm thick sapphire crystal by 800~nm pulses and filtered by a short-pass filter at 750 nm. The UV pump and visible probe beams were focused onto the sample using an OAP mirror, and the reflected probe was sent to a photodiode for detection with a lock-in scheme. The angle of incidence of the probe is $\sim$10$^{\circ}$, while the pump was set at normal incidence. The temporal chirp dispersion of the probe was found to be smooth, fitted by a polynomial function, and corrected through analysis. The polarization of the probe was set to be perpendicular to the 400~nm pump pulse. In this set of measurements, a hard-coated bandpass filter centered at 400~nm with full-width half maximum of 40~nm was used to ensure considerable photon flux. Therefore, this configuration yields a shorter pulse duration compared to the time-resolved HHG experiments. All the transient reflectivity spectroscopy measurements were also performed at 295~K.

\section{Time-dependent exact diagonalization simulations}
\label{sec:tded}
We use time-dependent exact diagonalization to solve the time evolution of the wavefunction of a single-band one-dimensional Fermi-Hubbard model, with the pump field entering the Hamiltonian via a Peierls' substitution:
\begin{align}
    H(t)=&-t_h\sum_{j\sigma}[e^{-iA(t)}c_{j\sigma}^\dagger c_{j+1\sigma}+e^{iA(t)}c_{j+1\sigma}^\dagger c_{j\sigma}] \nonumber\\&+ U\sum_{j}n_{j\uparrow}n_{j\downarrow}.
\end{align}
Here, $c_{j\sigma}^\dagger$ ($c_{j\sigma}$) creates (annihilates) an electron at site $j$ with spin $\sigma$. The parameter $t_h$ is the nearest-neighbor hopping integral and $U$ is the on-site Coulomb repulsion. $U=8t_h$ is set to simulate the strong correlations in La$_2$CuO$_4$ \cite{pavarini2001band,yoshida2006systematic}. We choose a chain size $L=12$ and a coarse-grained time set $\delta t=0.05t_h^{-1}$.

To compute the harmonic emission, we first use the electric current operator defined as
\begin{equation}
    J(t) = -it_h\sum_{j\sigma}[e^{-iA(t)}c_{j\sigma}^\dagger c_{j+1\sigma}-e^{iA(t)}c_{j+1\sigma}^\dagger c_{j\sigma}]
\end{equation}
to calculate the time-dependent current. HHG spectrum from charge acceleration is then given by
\begin{equation}
    S(\omega) = |\int \partial_t \langle J(t)\rangle e^{-i\omega t}dt|^2.
\end{equation}
Convergence of the harmonic emission and of the time evolution of the relevant observables used in this work were confirmed by performing the calculations for $L = 14$ and 16, yielding essentially the same results as for $L = 12$ (Supplemental Material Sec.~\ref{sec:system size dependence} \cite{SI}).

To characterize the evolution of elementary excitations, we calculate time-dependent single-particle spectral function of the form \cite{freericks2009theoretical}
\begin{equation}
\begin{split}
    \mathcal{A}(k,\omega;t) = \frac{1}{2\pi i}\iint &\, e^{i\omega(t_1-t_2)} \\
    & \times g(t_1;t)\,g(t_2;t)\,G_k^<(t_1,t_2)\,dt_1\,dt_2
\end{split}
\end{equation}
where $g(t_0;t)$ describes the probe pulse envelope and is given by a Gaussian function centered at $t_0$ with standard deviation $\sigma_p$
\begin{equation}
    g(t_0;t) = \frac{1}{\sqrt{2\pi}\sigma_p}e^{-(t-t_0)^2/2\sigma_p^2}
\end{equation}
and $G_k^<(t_1,t_2)$ is the two-time nonequilibrium lesser Green's function given by
\begin{equation}
    G_k^<(t_1,t_2) = i\theta(t_2-t_1)\langle c_k^\dagger(t_2)c_k(t_1)\rangle.
\end{equation}
Due to the Fourier transform limitation and the need to highlight the dynamical properties, we choose a probe width $\sigma_p=2t_h^{-1}$ as a compromise between time and frequency resolutions.

To measure how strongly charge-density fluctuations are correlated at a momentum $k$, we computed the instantaneous charge structure factor
\begin{equation}
    N(k,t)=\langle\psi(t)|\rho_{-k}\rho_k|\psi(t)\rangle,
\end{equation}
where $\rho_k=\sum_{q,\sigma}c^\dagger_{k+q,\sigma}c_{q,\sigma}$ is the charge density operator. This quantity reveals the tendencies of charge ordering and spatial patterns of density fluctuations.

We utilize the analytical expression for the holon dispersion of the one-dimensional Hubbard model at half filling to perform semiclassical analysis of subcycle HHG dynamics. The dispersion is parameterized through rapidity $q$, and the momentum $p_{h,d}(q)$ and energy $\varepsilon_{h,d}(q)$ are given by
\begin{align}
    p_h(q)&=p_d(q)+\pi\nonumber\\&=\frac{\pi}{2}-q-2\int_0^\infty\frac{d\omega}{\omega}\frac{\mathcal{J}_0(\omega)\sin{(\omega\sin{q})}}{1+\exp{\frac{u\omega}{2}}}\label{eq:p},\\
    \varepsilon_h(q)&=\varepsilon_d(q)\nonumber\\
    &=\frac{U}{2}+2\cos{q}+2\int_0^\infty\frac{d\omega}{\omega}\frac{\mathcal{J}_1(\omega)\cos{(\omega\sin{q})}e^{-\frac{U\omega}{4}}}{\cosh{\frac{u\omega}{4}}}\label{eq:e},
\end{align}
where $u=U/t_h$ is the correlation strength and $\mathcal{J}_n$ is the $n$-th Bessel function of the first kind \cite{essler2005one}. The resulting band structure is shown in Fig.~\ref{sfig:tr hhg spectrum}A \cite{SI}. The semiclassical three-step model describes the recombination process with the following equations of motion \cite{vampa2015semiclassical}
\begin{align}
    \frac{dx}{dt} &= \frac{\partial\varepsilon_g}{\partial k},\label{semi1}\\
    k(t) &= k_0 - A(t) + A_0,\label{semi2}
\end{align}
where $x(t)$ is the relative distance between the electron and the hole and $\varepsilon_g=\varepsilon_d+\varepsilon_h$ is the momentum-dependent energy gap. $k_0$ is the momentum with the smallest gap at which the tunneling probability is the largest \cite{andrei2026subgap} and $A_0$ is the vector potential at the instant of electron-hole pair creation. The harmonic emission from recombination occurs at $x=0$ with a frequency that equals the energy of the electron-hole pair. Then, we perform a subcycle decomposition of the current $J(t)$ by wavelet analysis to extract the temporal emission amplitude at each frequency. Time-domain HHG is computed by continuous wavelet transform
\begin{equation}
    I(s,\tau)=\frac{1}{\sqrt{|s|}}\int_{-\infty}^{\infty}I(t)\psi^*(\frac{t-\tau}{s})dt
\end{equation}
where $I(t)=\partial_t \langle J(t)\rangle$ is the charge acceleration used to calculate the HHG spectrum and $\psi$ is the wavelet. We use a Morlet wavelet to perform the analysis
\begin{equation}
    \psi(u)=\pi^{-\frac{1}{4}}e^{i\omega_0u}e^{-\frac{u^2}{2}}
\end{equation}
where $\omega_0=7.5$ (the dimensionless Morlet central frequency) is adopted to ensure adequate joint time and frequency resolution.

\section{Semiclassical evaluation of finite excursion-length effects}
\label{sec:excursion}
We used the dispersion of the one-dimensional Hubbard model and semiclassical equations of motion defined in Eqs.~\eqref{eq:p}--\eqref{semi2} to examine the effect of a finite doublon-holon excursion length on the HHG cutoff. For each driving field and photon energy, trajectories were initialized at $x(t_i)=0$ over the sampled pair-creation times $t_i$, with $A_0=A(t_i)$, and the relative coordinate was propagated using Eqs.~\eqref{semi1} and \eqref{semi2}. Recombination times $t_r>t_i$ were identified from subsequent zero crossings of the relative coordinate, $x(t_r)=0$, and the photon energy associated with each event was assigned from the instantaneous pair energy $\varepsilon_g[k(t_r)]$.

For each recolliding trajectory, we recorded the largest separation reached before recombination,
\begin{equation}
    l_{\mathrm{exc}}(t_i,t_r)=\max_{t_i\leq t\leq t_r}|x(t)|.
\end{equation}
The unconstrained cutoff was defined as the largest recombination energy among all sampled trajectories. To model a finite scattering-limited excursion, the analysis was repeated while retaining only trajectories satisfying $l_{\mathrm{exc}}\leq l_{\mathrm{max}}$, so that
\begin{equation}
    E_{\mathrm{cut}}(l_{\mathrm{max}})
    =\max_{\substack{t_i,t_r\\x(t_r)=0\\l_{\mathrm{exc}}\leq l_{\mathrm{max}}}}
    \varepsilon_g[k(t_r)].
\end{equation}
For each driving photon energy, we also determined the minimum excursion length required to access the unconstrained cutoff. The procedure was repeated as a function of $l_{\mathrm{max}}$, driving photon energy, and field amplitude.

\section{Semiconductor Bloch equations simulations}
\label{sec:sbe simulations}
We simulated the pump-induced modification of HHG by solving semiconductor Bloch equations for an effective two-band representation of the one-dimensional Hubbard spectrum \cite{haug2009quantum,kira2011semiconductor}. In the Houston frame, with $k(t)=q+A(t)$ in the adopted units, the microscopic interband coherence $p_q(t)$ and pair occupation $n_q(t)$ were propagated according to
\begin{align}
    \dot p_q(t)
    &=
    -\left[
        \frac{1}{T_2(t)}
        +\frac{i}{\hbar}\varepsilon_g(k)
    \right]p_q(t)
    -\frac{i}{\hbar}d_0E(t)\left[1-2n_q(t)\right],
    \label{eq:sbe-p}\\
    \dot n_q(t)
    &=
    -\frac{2}{\hbar}
    \operatorname{Im}\!\left[d_0E(t)p_q^*(t)\right]
    -\frac{n_q(t)-\bar n_q(t)}{\tau_p(t)}.
    \label{eq:sbe-n}
\end{align}
Here, $T_2(t)$ is the interband dephasing time, $\tau_p(t)$ is the momentum-relaxation time, and $\bar n_q(t)$ is the target distribution for momentum relaxation. The interband dipole $d_0$ was treated as momentum independent. At each time step, the normalization and effective temperature of $\bar n_q(t)$ were chosen to conserve the total pair number and kinetic energy.

The Brillouin-zone-averaged pair population was calculated as
\begin{equation}
    n(t)=\int_{\mathrm{BZ}}\frac{dq}{2\pi}\,n_q(t),
    \label{eq:sbe-density}
\end{equation}
and was used to define the density-dependent relaxation rates
\begin{align}
    \frac{1}{T_2(t)}
    &=
    \frac{1}{T_2^{(0)}}+\alpha n(t),
    \label{eq:sbe-t2}\\
    \frac{1}{\tau_p(t)}
    &=
    \frac{1}{\tau_p^{(0)}}+\beta n(t),
    \label{eq:sbe-taup}
\end{align}
where $T_2^{(0)}$ and $\tau_p^{(0)}$ are the unpumped values and $\alpha$ and $\beta$ parameterize the density dependence.

The interband polarization and the corresponding interband current were evaluated as
\begin{align}
    P_{\mathrm{inter}}(t)
    &=
    \int_{\mathrm{BZ}}\frac{dq}{2\pi}\,
    2\operatorname{Re}\!\left[d_0^*p_q(t)\right],
    \label{eq:sbe-polarization}\\
    j_{\mathrm{inter}}(t)
    &=
    \partial_t P_{\mathrm{inter}}(t),
    \label{eq:sbe-jinter}
\end{align}
while the intraband current was calculated from
\begin{align}
    j_{\mathrm{intra}}(t)
    &=
    \int_{\mathrm{BZ}}\frac{dq}{2\pi}\,
    \Delta v[k(t)]\,n_q(t),
    \label{eq:sbe-jintra}\\
    \Delta v(k)
    &=
    \frac{1}{\hbar}\frac{\partial\varepsilon_g(k)}{\partial k}.
    \label{eq:sbe-velocity}
\end{align}
The total current, $j(t)=j_{\mathrm{inter}}(t)+j_{\mathrm{intra}}(t)$, was restricted to the IR-pulse window and used to calculate the harmonic spectrum.

We solved Eqs.~\eqref{eq:sbe-p} and \eqref{eq:sbe-n} under the combined pump and infrared fields over the selected pump fluences, driving photon energies, and harmonic orders. For each condition, the corresponding harmonic yield was integrated over its spectral window. The pulse parameters and numerical settings are provided in the Supplemental Material Sec.~\ref{sec:sbe} \cite{SI}.

\bibliography{refs}

\clearpage
\onecolumngrid

\makeatletter
\setcounter{section}{0}
\setcounter{subsection}{0}
\setcounter{subsubsection}{0}
\def\thesection{\arabic{section}}
\def\thesubsection{\Alph{subsection}}
\def\thesubsubsection{\arabic{subsubsection}}
\let\@sectioncntformat\relax
\let\@hangfrom@section\@hang@from
\def\@hangfroms@section#1#2{#1#2}
\@removefromreset{equation}{section}
\setcounter{equation}{0}
\def\theequation@prefix{S}
\setcounter{figure}{0}
\renewcommand{\thefigure}{S\arabic{figure}}
\setcounter{table}{0}
\renewcommand{\thetable}{S\arabic{table}}
\makeatother

\begin{center}
{\large\bfseries Supplemental Material:\\Dynamical many-body control of high harmonic generation in a Mott insulator\par}
\vspace{1.5em}
Mingyao Guo, Ryo Noguchi, Mohammad Muntasir Hassan, Steven J. Gomez Alvarado, Reina M. Gomez, Yao-Jui Chan, Yu-Chieh Wen, Stephen D. Wilson, Yao Wang, and David Hsieh
\end{center}
\vspace{1em}

\tableofcontents
\addtocontents{toc}{\protect\SItocShow}

\setlength{\parindent}{2em}
\setlength{\parskip}{0.5em}
\setlength{\abovedisplayskip}{1.75em}
\setlength{\belowdisplayskip}{1.75em}
\setlength{\intextsep}{20pt plus 1.0pt minus 2.0pt}

\clearpage
\section{Material growth and characterization}
\label{sec:materials}
We employed a standard traveling solvent floating zone crystal growth recipe, using a flux with a molar composition of 4:1 CuO:La$_2$O$_3$ and a polycrystalline feed rod which contained 2\% excess moles of CuO to compensate for volatility and sustain the flux composition during the growth. Both the polycrystalline feed rod and the flux pellet were prepared via a conventional solid state reaction. Stoichiometric amounts of La$_2$O$_3$ (99.99\%, Alfa Aesar) and CuO (99.9995\% Thermo Scientific) powders were first ground into a homogeneous mixture. Prior to firing, powder was packed into a 5~mm diameter latex balloon and pressed under a hydrostatic pressure of 3275~bar in a cold isostatic press. Material for the feed rod underwent a three-step firing process with intermittent grinding and cold-pressing between each step to improve homogeneity: (1) 800~$^\circ$C for 24~hours in air, (2) 900~$^\circ$C for 24~hours in air, and (3) 1200~$^\circ$C for 12~hours in flowing O$_2$. Material for the flux underwent a similar three-step firing process: (1) 800~$^\circ$C for 24 hours in air, (2) 800~$^\circ$C for 24~hours in air, and (3) 850~$^\circ$C for 12~hours in air. The flux rod was then cut into a disc and placed atop a single crystalline seed of La$_2$CuO$_4$ for floating zone crystal growth in a four-mirror optical floating zone furnace with 300~W halogen lamps (Crystal Systems Corp. Model FZ-T-10000-H-VI-VPO-I-HR-PC). A growth rate of 1~mm/hr was maintained over several days under 1~atm dynamic pressure of flowing O$_2$.

\begin{figure*}[htb]
	\centering
	\includegraphics{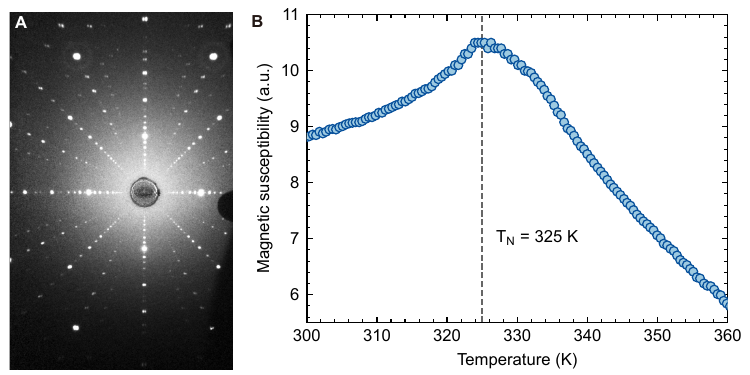}
	\caption{\textbf{Characterization of single crystalline La$_2$CuO$_4$.}
	(\textbf{A}) Laue X-ray diffraction pattern obtained from the polished (001) surface of the crystal used in this study. Shadows near the right of the image are from the sample holder. (\textbf{B}) Magnetic susceptibility of bulk La$_2$CuO$_4$ in arbitrary units. The dashed line denotes the deduced N\'eel temperature of 325 K. A field of 1500~Oe was applied perpendicular to the $c$ axis.}
	\label{sfig:sample characterization}
\end{figure*}

Typically, the incorporation of a small amount of interstitial oxygen during the crystal growth leads to as-grown crystals which are lightly hole-doped and superconducting. In order to remove the excess oxygen, the following vacuum annealing procedure was performed: heat from 30~$^\circ$C to 910~$^\circ$C over 3~hours, hold for 1~hour, cool to 600~$^\circ$C over 12~hours, cool to 500~$^\circ$C over 200~hours, and finally cool to room temperature over 12~hours \cite{PhysRevB.104.014304}. The resulting crystal was oriented using a Laue X-ray diffractometer and then diced/polished to expose the (001) crystallographic plane (Fig.~\ref{sfig:sample characterization}A). Magnetic susceptibility measurements were then carried out on a small piece from the annealed crystal using a Quantum Design Magnetic Property Measurement System (MPMS3) under an applied field of 1500~Oe perpendicular to the $c$ axis. Our magnetization measurements confirm the absence of a low-temperature diamagnetic signal and show a cusp at the AFM ordering temperature $T_\mathrm{N}=325~\mathrm{K}$ expected for the parent Mott insulating state of La$_2$CuO$_4$ \cite{keimer1992neel,PhysRevB.104.014304}, verifying the successful removal of excess interstitial oxygen (Fig.~\ref{sfig:sample characterization}B).

\clearpage

\section{Experimental setups}
\label{sec:optical setup}
\begin{figure*}[htb]
	\centering
	\includegraphics[width=0.9\textwidth]{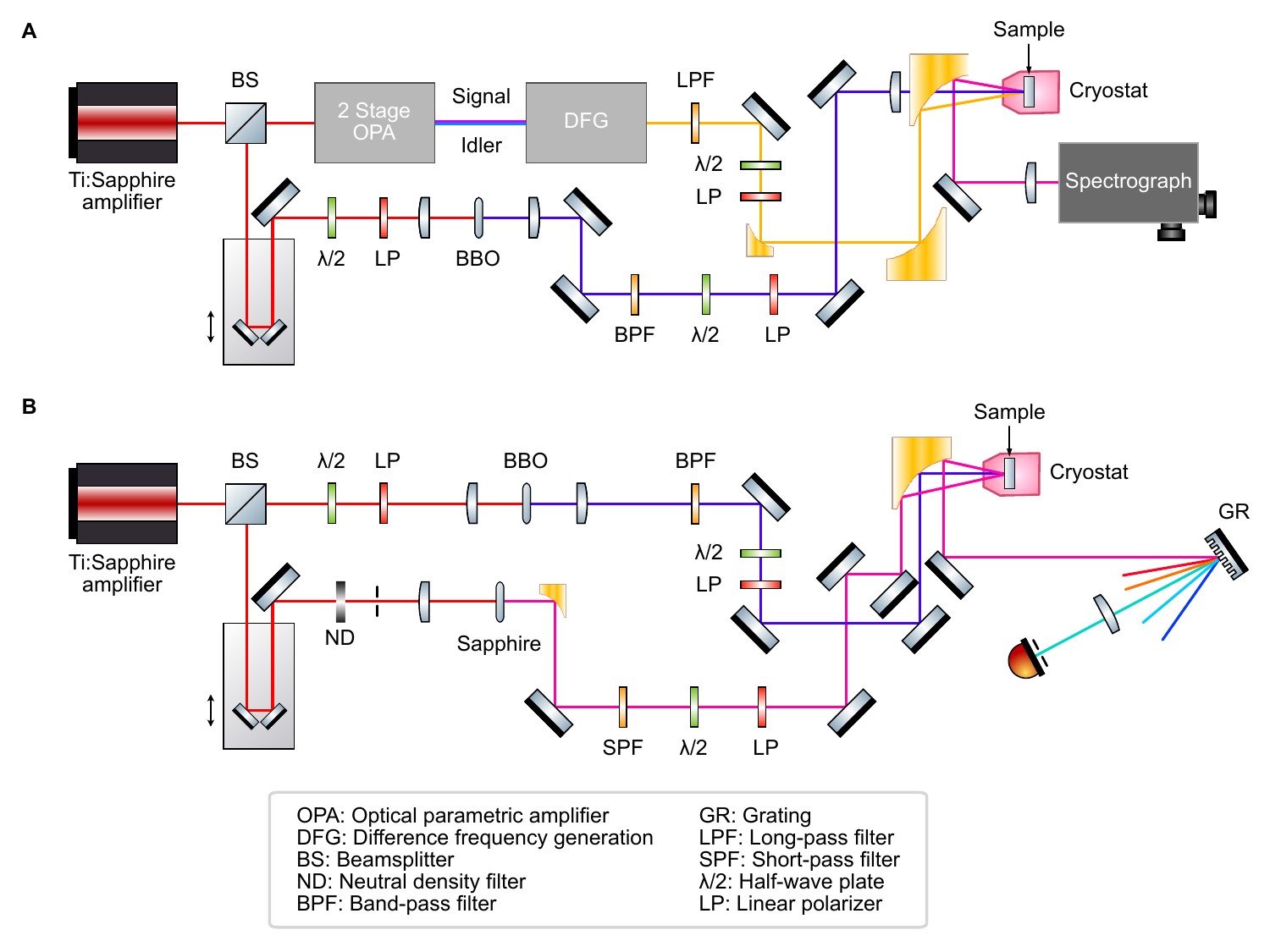}
	\caption{Schematic layout of the experimental (\textbf{A}) time-resolved HHG, and (\textbf{B}) transient reflectivity spectroscopy setups.}
	\label{sfig:setup}
\end{figure*}

\begin{figure*}[htb]
	\centering
	\includegraphics[width=0.8\textwidth]{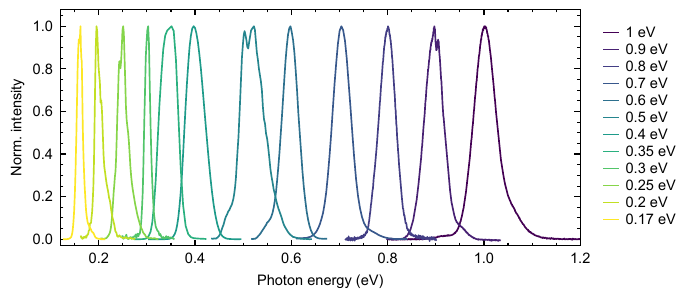}
	\caption{\textbf{Characterization of the IR driving pulses for HHG.}
	The spectra of various IR pulses used in the experiment.}
	\label{sfig:pulse characterization}
\end{figure*}

The time-resolved HHG setup (Fig.~\ref{sfig:setup}A) is based on a femtosecond titanium-sapphire laser amplifier (repetition rate, 1~kHz; pulse duration, 35~fs; center wavelength, 800~nm), which is used to pump a dual-stage optical parametric amplifier (OPA). The output wavelength of signal/idler pulses can be readily tuned between 1.17 $\mu$m and 2.5~$\mu$m (signal/idler pulse duration, 50/70~fs). Mid-IR pulses are generated by their difference frequency mixing in an AgGaS$_2$ (2.6$\sim$4.5~$\mu$m) or GaSe (4.5$\sim$10~$\mu$m) crystal in a noncollinear geometry (pulse duration, $\sim$100~fs). Each IR pulse wavelength was characterized before the measurements using a home-built grating-based spectrometer (Fig.~\ref{sfig:pulse characterization}). For mid-IR excitations, a germanium long-pass filter was employed to filter out the residual near-IR pulses. 

High harmonics (HHs) were obtained by focusing the IR beams obliquely (incident angle, 10$^{\circ}$) onto the sample by an off-axis parabolic (OAP) mirror with a 1/$e^2$ beam diameter on the order of 60~$\mu$m depending on the exact excitation wavelength. The highest employed peak driving field of 31~MV/cm is close to the observed long-time damage threshold of La$_2$CuO$_4$. The resulting reflected spectrum was collected by the same OAP mirror and recorded by a home-built grating-based spectrometer with an indium gallium arsenide detector array (1$\sim$1.7~$\mu$m) and a cooled silicon charge-coupled device (0.2$\sim$1.1~$\mu$m), respectively. All spectra are corrected for the reflectance of mirrors, the diffraction efficiency of the grating, and the quantum efficiency of the detector.

To generate the UV pump excitation, the fundamental 800~nm pulses were focused onto a 0.2~mm thick type-I $\beta$-barium borate crystal, yielding second harmonic with central wavelength at 400~nm. A hard-coated bandpass filter centered at this wavelength (full-width half maximum, 10~nm) was used to filter out the residual pump pulses. The 400-nm beam was focused onto the sample at normal incidence through a hole in the OAP mirror by a lens with a 1/$e^2$ beam diameter of 150~$\mu$m. A shortpass dichroic beamsplitter (edge wavelength, 414~nm) was employed to filter out the scattered pump photons.

The IR pulses were linearly polarized along the $a$ axis of the $c$-axis oriented bulk La$_2$CuO$_4$ crystal (Sec. \ref{sec:orientation}). The UV pump pulse was linearly polarized along the $b$ axis to minimize nonlinear frequency-mixing signals with the drive field during their temporal overlap. All the HHG measurements were carried out at 295 K.

The broadband visible pulses for transient reflectivity spectroscopy (Fig.~\ref{sfig:setup}B) were produced by supercontinuum generation in a 4-mm thick sapphire crystal pumped by 800~nm pulses and filtered by a short-pass filter at 750~nm. The UV pump and visible probe beams were focused onto the sample using an OAP mirror, and the reflected probe was sent to a photodiode for detection with a lock-in scheme. The angle of incidence of the probe is $\sim$10$^{\circ}$, while the pump was set at normal incidence. The temporal chirp dispersion of the probe was found to be smooth, fitted by a polynomial function, and corrected through analysis. The polarization of the probe was set to be perpendicular to the 400~nm pump pulse. In this set of measurements, a hard-coated bandpass filter centered at 400~nm with full-width half maximum of 40~nm was used to ensure considerable photon flux. Therefore, this configuration yields a shorter pulse duration compared to the time-resolved HHG experiments. All the transient reflectivity spectroscopy measurements were also performed at 295 K.

\clearpage
\section{Crystallographic orientation dependence of HHG}
\label{sec:orientation}
\begin{figure*}[htb]
	\centering
	\includegraphics[width=\textwidth]{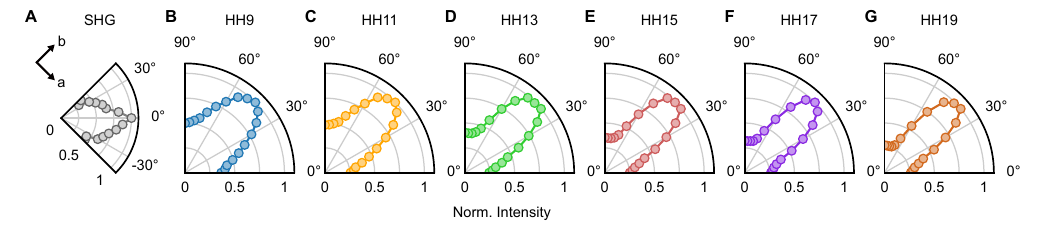}
	\caption{\textbf{Crystal orientation dependence of SHG and HHG from La$_2$CuO$_4$.}
	(\textbf{A}) RA-SHG pattern acquired in the 'S$_{\rm{in}}$-P$_{\rm{out}}$' geometry. The inset shows the deduced crystal axes. (\textbf{B})-(\textbf{G}) Anisotropic 9th to 19th order HHG of an 8450~nm laser.}
	\label{sfig:orientation dependence}
\end{figure*}

Crystal orientation was determined by a high-speed rotational-anisotropy second harmonic generation (RA-SHG) setup \cite{harter2015high}. An 800-nm laser was used as the probe and the SHG signal was collected in the 'S$_{\rm{in}}$-P$_{\rm{out}}$' geometry (Fig.~\ref{sfig:orientation dependence}A). Based on the \emph{mmm} point group symmetry of the crystal in its low-temperature orthorhombic phase, we assign the \emph{a} and \emph{b} axes shown in the inset of Fig.~\ref{sfig:orientation dependence}A. Because the in-plane anisotropy along the \emph{a} and \emph{b} axes is negligible \cite{PhysRevB.104.014304}, we refer to them as the Cu-O-Cu bond directions without distinguishing them here.

HH responses were subsequently recorded on the same sample once its orientation was aligned. Using a mid-IR laser of 8450~nm, we collected HHG from the 9th order to the 19th order as a function of linear polarization angle (Fig.~\ref{sfig:orientation dependence}, B to G). All the harmonic orders exhibit distinct anisotropy following crystallographic axes, with the emission maximizing when the driving field was polarized along the \emph{a} (\emph{b}) axis. This behavior is consistent with the CT nature of the lowest-energy interband transition. Minor signal asymmetry with respect to the 45$^\circ$ axis may come from an oblique-incident reflection geometry of the experimental setup (Fig.~\ref{sfig:setup}A) and the slightly different efficiency of S- and P-polarized light of the diffraction grating inside the spectrometer. 

\clearpage

\section{Wavelength-dependent HHG}
\begin{figure}[htb]
	\centering
	\includegraphics[width=\textwidth]{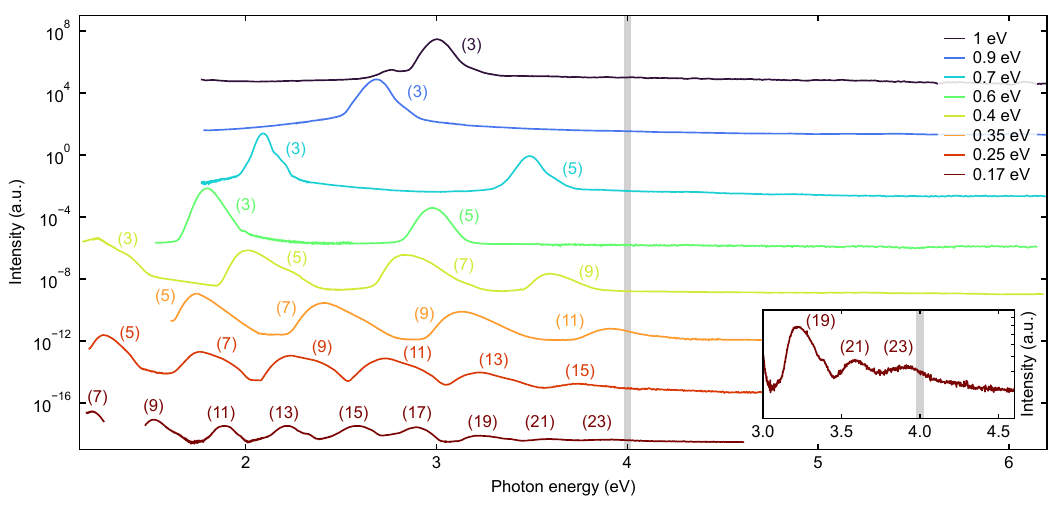}
	\caption{\textbf{Wavelength dependence of HHG from La$_2$CuO$_4$.}
	Harmonic spectra generated by various driving wavelengths. The inset provides a magnified view of UV emission under 0.17~eV drive. Curves are vertically offset for clarity.}
	\label{sfig:wavelength dependence}
\end{figure}

To further illustrate the $\sim4$~eV cutoff discussed in the main text, we show supplementary HHG spectra driven across a broader range of wavelengths (Fig.~\ref{sfig:wavelength dependence}). The peak field strength is kept around 30~MV/cm to preclude a trivial origin of the universal cutoff from insufficient field strength (Fig.~\ref{fig:m1}, D and E). In the near-IR regime, third harmonic generation (THG) dominates the spectrum and even appears visible to the eye. However, fifth harmonic generation (FHG) only emerges when its photon energy falls below $\sim$4 eV. Similarly, mid-IR driving generates a plateau of higher order harmonics extending into the visible and UV regimes, characteristic of non-perturbative HHG, but they are absent beyond this energy. In addition, for short wavelengths, SHG remains unresolvable in the emission spectrum, consistent with a recent report on another parent cuprate Sr$_2$CuO$_2$Cl$_2$ \cite{torre2021mirror}.

To provide more evidence for the existence of a global cutoff energy, we performed fine-step near-IR wavelength scans where HH energies approach 4~eV. Because $\sim$1550~nm (where FHG is 4~eV) lies near the tuning edge of our OPA pumped by the titanium-sapphire laser, we utilized a separate laser system only for this specific set of measurements. The near-IR laser pulses were generated by a dual-stage OPA (Light Conversion Inc., ORPHEUS-F; pulse duration, sub-100~fs) pumped by 1030-nm pulses generated by a ytterbium-doped
potassium gadolinium tungstate amplifier (Light Conversion Inc., PHAROS; repetition rate, 2~kHz; pulse duration, 325~fs). The experimental geometry remained identical, and HHG spectra were recorded by an Andor Shamrock SR-500i imaging spectrometer.

\begin{figure*}[htb]
	\centering
	\includegraphics[width=0.9\textwidth]{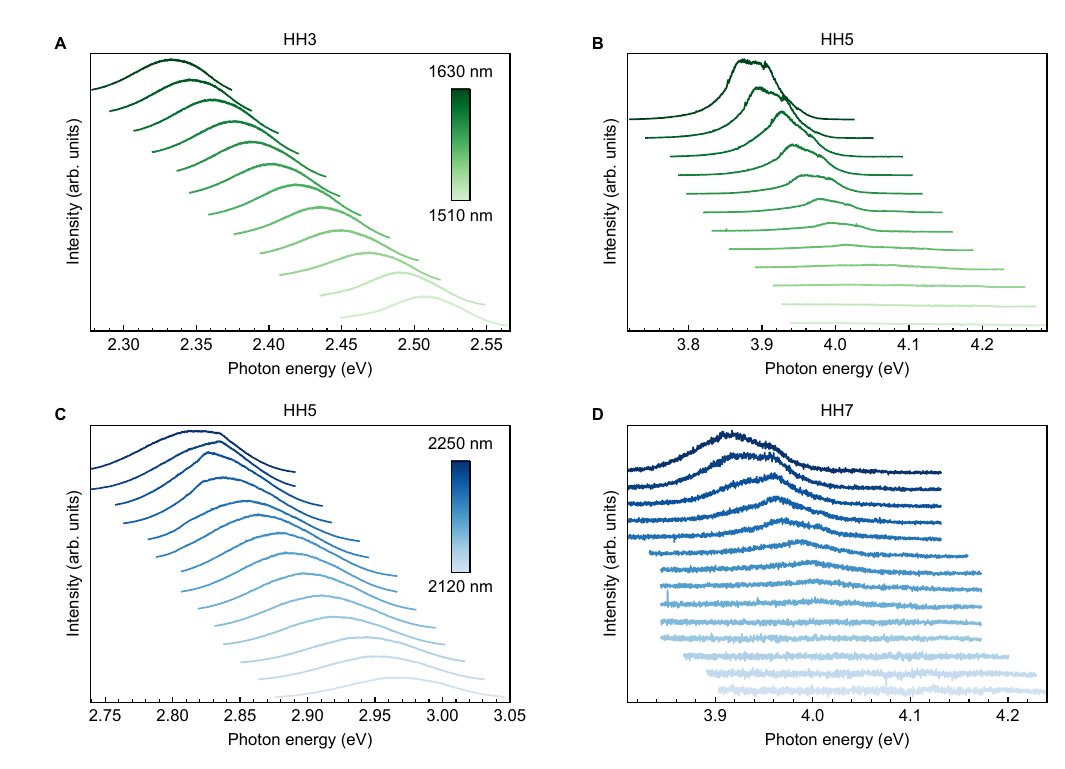}
	\caption{\textbf{Detailed wavelength dependence of HHG from La$_2$CuO$_4$ close to the cutoff near 4~eV.}
	(\textbf{A})-(\textbf{B}) THG and FHG spectra as the driving wavelength is tuned from 1510~nm to 1630~nm (bottom to top). (\textbf{C})-(\textbf{D}) FHG and seventh order harmonic spectra as the driving wavelength is tuned from 2120~nm to 2250~nm (bottom to top). The curves in each panel are vertically offset for clarity.}
	\label{sfig:wavelength dependence dense}
\end{figure*}

Figure~\ref{sfig:wavelength dependence dense}A and B show the THG and FHG spectra of drive wavelength from 1630~nm to 1510~nm. While the THG intensity remains relatively comparable in this narrow spectral range (Fig.~\ref{sfig:wavelength dependence dense}A), FHG yield diminishes drastically (Fig.~\ref{sfig:wavelength dependence dense}B) as the driving wavelength shortens. This trend directly opposes the standard observations and expectations in the absence of a band structure cutoff \cite{liu2018wavelength}. Instead, it demonstrates that the abrupt disappearance of FHG beyond $\sim$4~eV is enforced by the energy constraint of the correlated electronic band structure: a momentum-dependent gap prevents doublon-holon recombination beyond this threshold. We were unable to detect any seventh order harmonic emission under the same conditions, suggesting a vanishing tunneling probability from the ZRS band to higher-lying NB oxygen bands. The conclusion is corroborated by complementary measurements with laser wavelength from 2250~nm to 2120~nm, where the FHG intensity remains stable (Fig.~\ref{sfig:wavelength dependence dense}C) but the seventh harmonic becomes undetectable in our setup once its energy exceeds $\sim$4~eV (Fig.~\ref{sfig:wavelength dependence dense}D). 

While earlier time-dependent Schr\"{o}dinger equation simulations of wavelength scaling of the cutoff energy in solid-state HHG show that intraband emission yields a constant cutoff while interband radiation is associated with a linear cutoff with the wavelength \cite{liu2017wavelength}, we stress that other features in the model fail to capture our experimental results. Specifically: (i) our observed cutoff stops increasing after saturation at $\sim4$~eV even if we use stronger electric fields, while it grows linearly in this model; (ii) the linear dependence of the interband HHG cutoff on the wavelength requires that the cutoff energy has not yet reached the maximal conduction-to-valence band separation, otherwise it will saturate at this energy. Consequently, these data provide clear evidence of the importance of electronic structure in HHG from La$_2$CuO$_4$ based on carrier recombination.

\clearpage

\section{Semiclassical evaluation of finite excursion length effects on the solid-state HHG cutoff}
\label{sec:finite excursion length}
A recent study attributed a constant HHG cutoff energy in liquids to a finite electron excursion length limited by scattering \cite{mondal2023high}. Here, we evaluate the applicability of this mechanism to our results in La$_2$CuO$_4$.

Firstly, we review the key idea of the semiclassical description of HHG and cutoff scaling including the scattering effects proposed in Ref.~\cite{mondal2023high}. An electron driven by a one-dimensional electric field $E(t)=E_0\cos{\omega t}$ ($\omega$ is the laser angular frequency) and ionized at time $t_i$ follows the trajectory
\begin{equation}
    x(t) = \frac{eE_0}{m\omega^2}[\cos{\omega t_i} - \cos{\omega t} - \omega(t-t_i)\sin{\omega t_i}].\label{rec}
\end{equation}
The time $t_r$ at which the electron returns to the parent ion is given by the condition $x(t_r) = 0$. During the electron motion, the maximum excursion occurs at $t^*$ when $\frac{dx}{dt}=0$, that is,
\begin{equation}
    \frac{eE}{m\omega}(\sin{\omega t^*} - \sin{\omega t_i}) = 0,
\end{equation}
which has a simple analytical solution $\omega t^* = \pi - \omega t_i$. The maximal excursion length $l_\mathrm{exc}(t_i)$ is therefore
\begin{equation}
    l_\mathrm{exc}(t_i) = x(t^*) = \frac{eE}{m\omega^2}[2\cos{\omega t_i} - (\pi - 2\omega t_i)\sin{\omega t_i}].
\end{equation}
If scattering restricts the electron excursion length to a finite distance $l_\mathrm{max}<\mathrm{max}_{t_i} l_\mathrm{exc}(t_i)\approx1.145\frac{eE}{m\omega^2}$, the most energetic trajectories are forbidden and the cutoff will be reduced. In this regime, the cutoff energy corresponds to a trajectory that satisfies $l_\mathrm{exc}=l_\mathrm{max}$. Namely, the modified ionization time $t_i^\prime$ is determined by the solution of
\begin{equation}
    2\cos{\omega t_i^\prime} - (\pi - 2\omega t_i^\prime)\sin{\omega t_i^\prime} = \frac{m\omega^2 l_\mathrm{max}}{eE} = L_\mathrm{red}.
\end{equation}
The recombination time $t_r^\prime$ can then be solved using Eq.~\eqref{rec}. Numerically, it is found that when $L_\mathrm{red}$ is not too large (in the low frequency and high field regime), the accumulated kinetic energy reduces to a wavelength-independent, field linear form
\begin{equation}
    \Delta E_K = \frac{e^2 E_0^2}{2m\omega^2}(\sin{\omega t_r^\prime} - \sin{\omega t_i^\prime})^2\approx\frac{3.73}{4}eEl_\mathrm{max}.
\end{equation}

To test whether this liquid-phase scattering mechanism applies to solid-state band structures, we present a toy-model numerical simulation of a one-dimensional half-filled Hubbard model appropriate for cuprates. The dispersion is defined by Eqs.~\eqref{eq:1d hubbard p} and \eqref{eq:1d hubbard e} in Sec.~\ref{sec:subcycle analysis}, as shown in Fig.~\ref{sfig:tr hhg spectrum}A. The dynamics is then given by
\begin{align}
    \frac{dx}{dt} &= \frac{\partial\varepsilon_g}{\partial k},\label{si:semi1}\\
    k(t) &= k_0 - A(t) + A_0,\label{si:semi2}
\end{align}
where $x(t)$ now is the relative distance between the electron and the hole and $\varepsilon_g=\varepsilon_c-\varepsilon_v$ is the momentum-dependent energy gap. Assuming an external drive $A(t)=E_0/\omega\cos{\omega t}$, HHG emits when $x(t_r)=0$ and its frequency is $\varepsilon_g(k(t_r))$.

\begin{figure*}[htb]
	\centering
	\includegraphics[width=\textwidth]{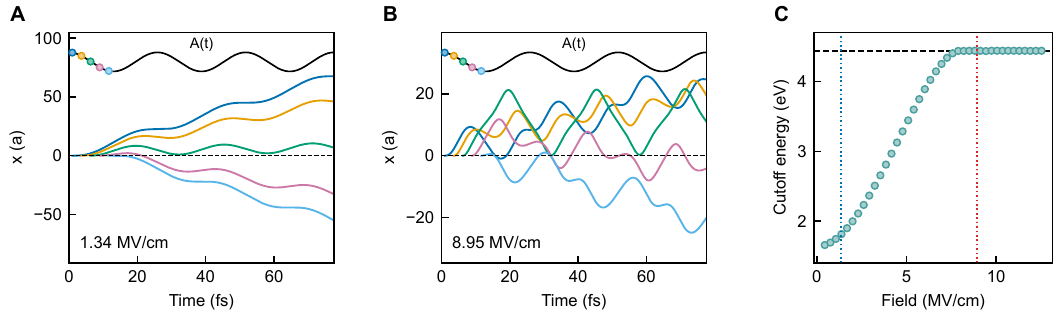}
	\caption{\textbf{Trajectory analysis and cutoff of HHG in the 1D Hubbard model without limiting excursion length.}
	(\textbf{A}) Relative distance $x(t)$ of the electron and the hole following their creation at different times indicated by the dots on top of the vector potential trace shown as a black line at the top of the panel at low field strength $F=1.34$~MV/cm. (\textbf{B}) Same as (\textbf{A}) but with a high field strength $F=8.95$~MV/cm. (\textbf{C}) The cutoff energy as a function of the driving field. The black dashed line denotes the maximal gap between the two Hubbard bands. The blue and red dotted lines mark the field strengths used in (\textbf{A}) and (\textbf{B}), respectively. The driving photon energy employed in this simulation is $\hbar\omega=0.16$~eV.}
	\label{sfig:dynamics wo limiting exc}
\end{figure*}

Let us start without limiting the electron excursion length. Figure~\ref{sfig:dynamics wo limiting exc}A and B show the relative distance $x(t)$ of the electron and the hole following their creation at different times with a weak (1.34~MV/cm) and strong (8.95~MV/cm) field, respectively. It can be clearly seen that electron-hole pairs undergo dynamical Bloch oscillations under the intense AC drive and only those created at certain times can recombine. Remarkably, the cutoff energy scales quasi-linearly with the electric field until it saturates at the maximal separation between the two Hubbard bands (Fig.~\ref{sfig:dynamics wo limiting exc}C), consistent with previous reports \cite{vampa2015semiclassical}.

\begin{figure}[htb]
	\centering
	\includegraphics{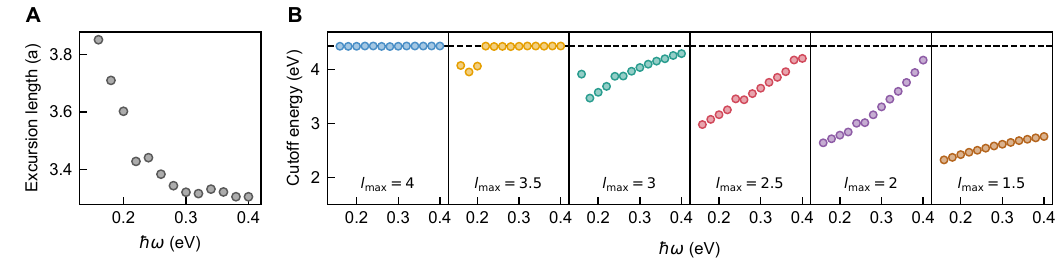}
	\caption{\textbf{Effect of finite excursion length on HHG in the 1D Hubbard model.}
	(\textbf{A}) Relationship between the minimal excursion length (in units of the lattice constant $a$) to achieve the highest cutoff and the driving photon energy. (\textbf{B}) The cutoff energy as a function of the driving photon energy in the presence of a finite excursion length $l_\mathrm{max}$ induced by scattering. The black dashed line denotes the maximal gap between the two Hubbard bands. The field strength employed in this simulation is $F=22.4$~MV/cm.}
	\label{sfig:dynamics w limiting exc}
\end{figure}

We then exert a constraint on the maximal excursion length to mimic the electron scattering effects. Figure~\ref{sfig:dynamics w limiting exc}A shows the minimal excursion length to reach the highest emission energy as a function of the driving photon energy in the strong-field regime ($F=22.4$~MV/cm). Note that here the HHG cutoff reaches the plateau under all driving photon energies between 0.16 and 0.4~eV, and the minimal excursion length decreases from $3.85a$ to $3.3a$ across this range. The wiggles here come from the fact that the recombining trajectories form discrete families distinguished by the number of driving half-cycles elapsed between pair creation and recombination, and while the minimal excursion of each family evolves continuously with the photon energy, the identity of the family that realizes the global minimum switches at particular photon energies, producing the kinks and small steps visible in Fig.~\ref{sfig:dynamics w limiting exc}A. When $l_\mathrm{max}=3.5$, as seen in Fig.~\ref{sfig:dynamics w limiting exc}A, only the lowest photon energies ($\hbar\omega\leq0.2$~eV) are affected, and hence their cutoff energies become lower than the rest as a result of the paths that lead to the highest HHG energy being eliminated (Fig.~\ref{sfig:dynamics w limiting exc}B). As $l_\mathrm{max}$ becomes smaller, more electron-hole paths turn ineffective, and the cutoff in the low-energy regime becomes lower. The same trajectory-branch switching underlies the discontinuities in Fig.~\ref{sfig:dynamics w limiting exc}B: the constrained cutoff is set by the most energetic trajectory whose excursion stays below $l_\mathrm{max}$, so whenever a trajectory family enters or leaves the allowed set as the photon energy is tuned, the cutoff jumps abruptly between the values supported by neighboring families. Such non-smooth structures are intrinsic to the semiclassical trajectory analysis and in a real material they would be smeared out by dephasing and by the statistical distribution of scattering lengths. When $l_\mathrm{max}<3.3$, the plateau regime in Fig.~\ref{sfig:dynamics wo limiting exc}C becomes out of reach for all photon energies (Fig.~\ref{sfig:dynamics w limiting exc}B), and the cutoff simply grows with increasing photon energies.

To summarize, in all circumstances, a wavelength-independent HHG cutoff seen in liquids induced by electron scattering effects is not reproduced by our simulation in a solid with the holon dispersion of a half-filled one-dimensional Hubbard model. We note that within the semiclassical analysis it only appears when (i) the electric field is strong enough to drive electrons and holes to the Brillouin zone boundary and recombine without any limitations on their excursion length, or (ii) the field is still strong enough and the excursion length is finite but not too short ($l_\mathrm{max}>3.85a$ in the case of Fig.~\ref{sfig:dynamics w limiting exc}), which effectively means that there is no constraint on dynamics that leads to the highest harmonic energy and the cutoff still saturates at the maximum of $\varepsilon_g(k)$. Consequently, we rule out the finite excursion length of electrons and holes imposed by scattering as the origin of the global cutoff energy in HHG spectra from La$_2$CuO$_4$.

\clearpage

\section{Electric-field-dependent HHG}
\begin{figure}[hb]
	\centering
	\includegraphics[width=\textwidth]{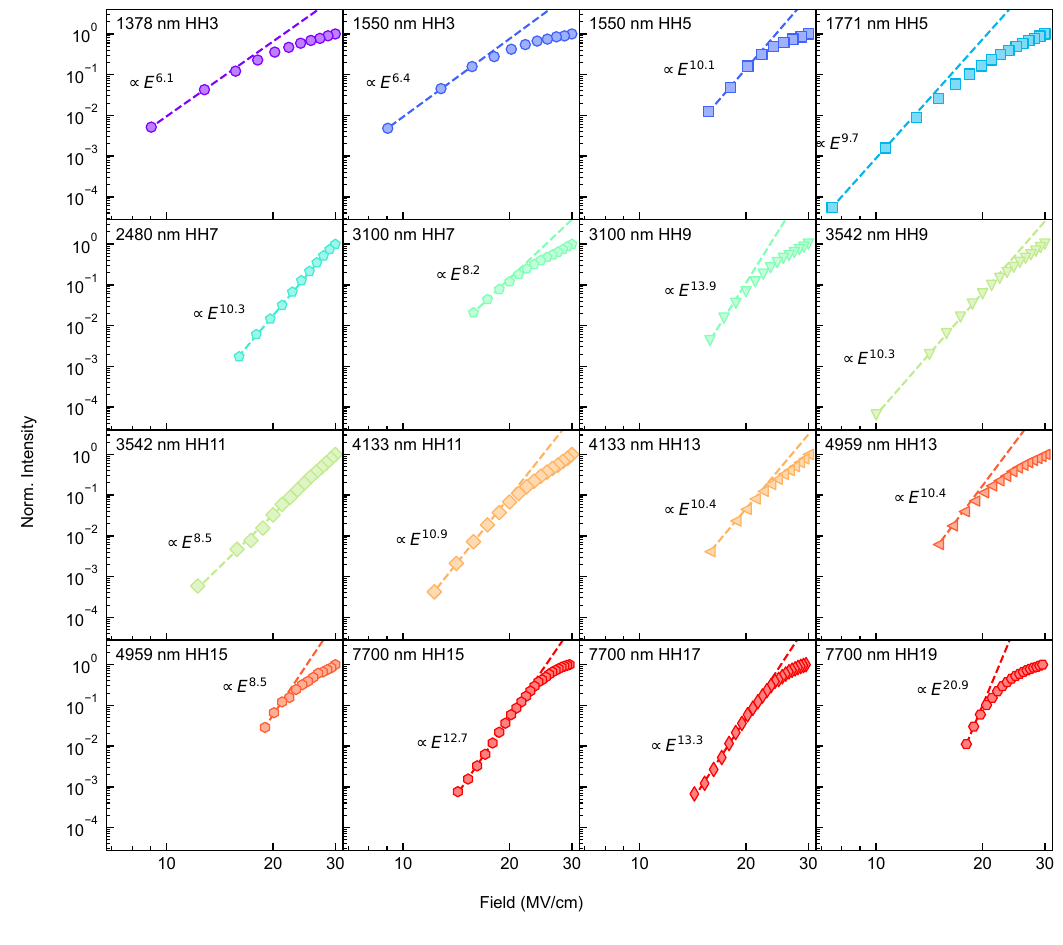}
	\caption{\textbf{Electric field dependence of HHG from La$_2$CuO$_4$.}
	Normalized integrated HHG intensities as a function of peak field strength across a range of driving wavelengths spanning from near-IR to mid-IR on a log-log scale. The dashed line in each panel represents power-law fit of the HH yields extracted at low fields.}
	\label{sfig:field dependence}
\end{figure}

We show supplementary data on electric-field dependence of HHG at various driving conditions to confirm its highly non-perturbative nature. The responses can be classified into two regimes (Fig.~\ref{sfig:field dependence}). For short wavelengths and low harmonic orders, the emission initially follows perturbative scaling before deviating from the corresponding power-law functional forms in the strong-field regime (Fig.~\ref{sfig:field dependence}, first row). On the other hand, for long wavelengths and high harmonic orders, the emission at low driving field already violates the perturbative nonlinear optical scaling, and ultimately saturates at high fields, again beyond the description of Kerr-type harmonics (Fig.~\ref{sfig:field dependence}, second to fourth rows). In HHG experiments presented in the main text, the peak field was maintained firmly within the non-perturbative regime.

\clearpage

\section{Dynamics of time-resolved HHG spectroscopy}
\begin{figure*}[htb]
	\centering
	\includegraphics[width=\textwidth]{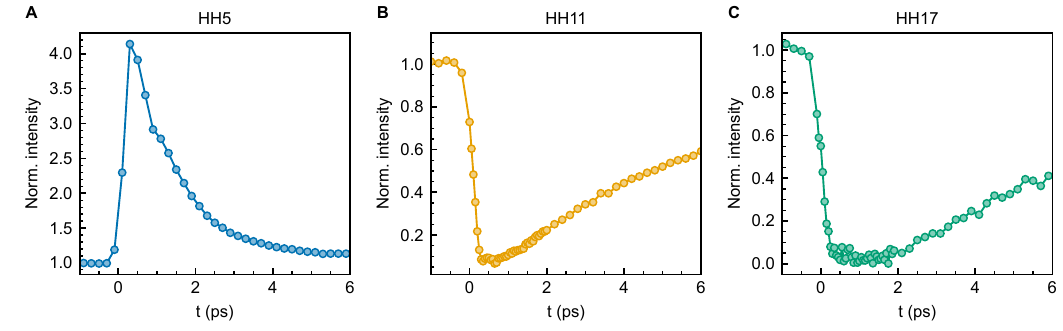}
	\caption{\textbf{HHG transients from La$_2$CuO$_4$ under 7700~nm driving field.}
	(\textbf{A})-(\textbf{C}) Time-resolved intensity of the 5th, 11th, and 17th harmonic order driven by a 25.4~MV/cm 7700~nm drive pulse, following excitation by a 2~mJ/cm$^2$ 400~nm pulse.}
	\label{sfig:hhg dynamics}
\end{figure*}

Representative time-resolved HHG traces are presented in Fig.~\ref{sfig:hhg dynamics}. All harmonic orders exhibit a pronounced response to the 400~nm pump pulse near zero time delay, followed by a subsequent decay back towards their equilibrium intensities. Because this study specifically investigates the role of nonthermal photocarriers rather than their associated lattice heating effects, the main text focuses on the harmonic spectra at a delay of 300~fs. At this specific delay, the nonequilibrium electronic response is maximal, yet the system remains non-thermalized. Crucially, a 300~fs delay ensures the temporal separation between the UV pump and IR drive pulses, thereby completely eliminating the artifacts from nonlinear optical wave-mixing \cite{zaks2012experimental,langer2016lightwave}.

\clearpage

\section{Driving electric field dependence of transient HHG}
\label{sec:giant intraband HHG}
\begin{figure}[hb]
	\centering
	\includegraphics{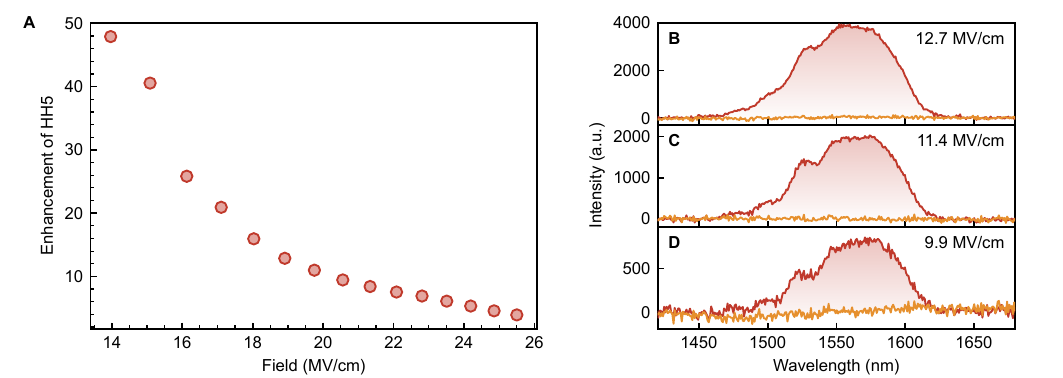}
	\caption{\textbf{Photo-induced enhancement of FHG from La$_2$CuO$_4$ under 7700~nm driving field.}
	(\textbf{A}) Relative enhancement of FHG upon photo-doping with a 2~mJ/cm$^2$ 400~nm pulse as a function of the peak driving field. (\textbf{B})-(\textbf{D}) FHG spectra before (orange) and immediately after (red) the same pump excitation condition as in (\textbf{A}) at different peak field strengths indicated in each panel. The spectra were recorded at a time delay of 300~fs.}
	\label{sfig:hh5 enhancement}
\end{figure}

To further substantiate the intraband origin of the below-gap harmonics and to highlight the potential for massive HHG amplification via photo-carrier injection, we evaluated the transient HHG enhancement as a function of the peak driving field. Figure~\ref{sfig:hh5 enhancement}A displays the dependence of enhanced FHG upon photo-doping on the 7700~nm driving field strength at a constant 400~nm pump fluence of 2~mJ/cm$^2$. The relative enhancement increases dramatically as the driving field decreases, whose amplitude can be more than one order of magnitude. Remarkably, when the drive field is too weak to generate observable harmonics in equilibrium, photo-doping could even stimulate the emergence of FHG out of equilibrium (Fig.~\ref{sfig:hh5 enhancement}, B-D), yielding an effectively divergent relative enhancement. 

This behavior directly reflects the intraband mechanism of below-gap HHG. Because the pump-injected carriers are mutually incoherent after creation, they cannot coherently contribute to the phase-matched recombination required for interband HHG. Nevertheless, intraband dynamical Bloch oscillations rely only on the individual coherence of doublons and holons. Consequently, these photo-carriers act as additional, independent sources aside from those created by the driving field itself for intraband HHs. In this measurement, by maintaining a constant pump fluence while weakening the driving field, the ratio of pump-injected carriers to drive-generated carriers increases, amplifying the relative harmonic yield.

Following standard reflection-geometry models for bulk nonlinear optics \cite{bloembergen1962light,vampa2018observation}, the back-reflected harmonic emerges coherently from a surface sheet of effective thickness $L_{\rm coh}^{(n)} = \pi/|\Delta k_{\rm refl}^{(n)}| = \lambda_{\rm harm}/{2[\eta(\omega)+\eta(n\omega)]}$ ($\eta$ is the complex sum of refractive index and extinction coefficient). For below-gap harmonics studied here, $L_{\rm coh}$ ranges from $\sim$162~nm at 1540~nm down to $\sim$105~nm at 1000~nm, which significantly exceeds the UV pump penetration depth $\delta_{\rm UV} = 61.5$~nm (see discussions in Sec.~\ref{sec:saturation model}). Therefore, the intrinsic photo-doped enhancement exceeds our measured ratio by a volume factor $L_{\rm coh}^{(n)}/\delta_{\rm UV}$. At the highest employed driving field for FHG of 7700~nm, this yields an estimated intrinsic enhancement of over one order of magnitude. However, we also note that these calculations rely on equilibrium optical constants that dynamically evolve during excitation, and thus this enhancement level only serves as an estimation.

More importantly, we expect this approach to generalize to other solids in which the intraband mechanism dominates HHG, e.g., SiO$_2$ \cite{luu2015extreme,garg2016multi} and GaSe \cite{schubert2014sub,hohenleutner2015real}. Since intraband HHG in solid targets has been found to exhibit robust waveforms against intensity and phase fluctuations of the driving field \cite{garg2018ultimate}, incorporating the photo-doping technique to existing technologies may pave the way to ultimately realizing compact solid-state extreme-UV attosecond sources by significantly enhancing their yields.

\begin{figure*}[htb]
	\centering
	\includegraphics[width=\textwidth]{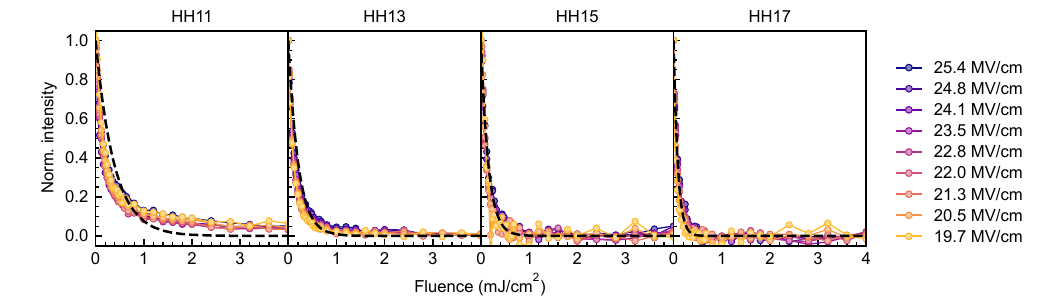}
	\caption{\textbf{Driving field dependence of time-resolved above-gap HHG from La$_2$CuO$_4$.}
	Normalized intensity of the 11th through the 17th harmonic orders as a function of the 400~nm pump fluence at different peak electric fields of the 7700~nm drive. The black dashed line in each panel indicates the fitting from a phenomenological saturation model discussed in Sec.~\ref{sec:saturation model}.}
	\label{sfig:above gap hhg field dep}
\end{figure*}

In contrast, above-gap transient HHG is largely insensitive to the driving field strength (Fig.~\ref{sfig:above gap hhg field dep}). Provided the light-matter interaction remains non-perturbative, this field-independent suppression aligns exceptionally well with our phenomenological dephasing model presented in Sec.~\ref{sec:saturation model} (black dashed lines), reinforcing the distinct interband nature of these high-energy harmonics.

\clearpage

\section{Fitting of the charge-transfer transition in reflectivity transients}
\label{fitting R}
\begin{figure*}[hb]
	\centering
    \includegraphics{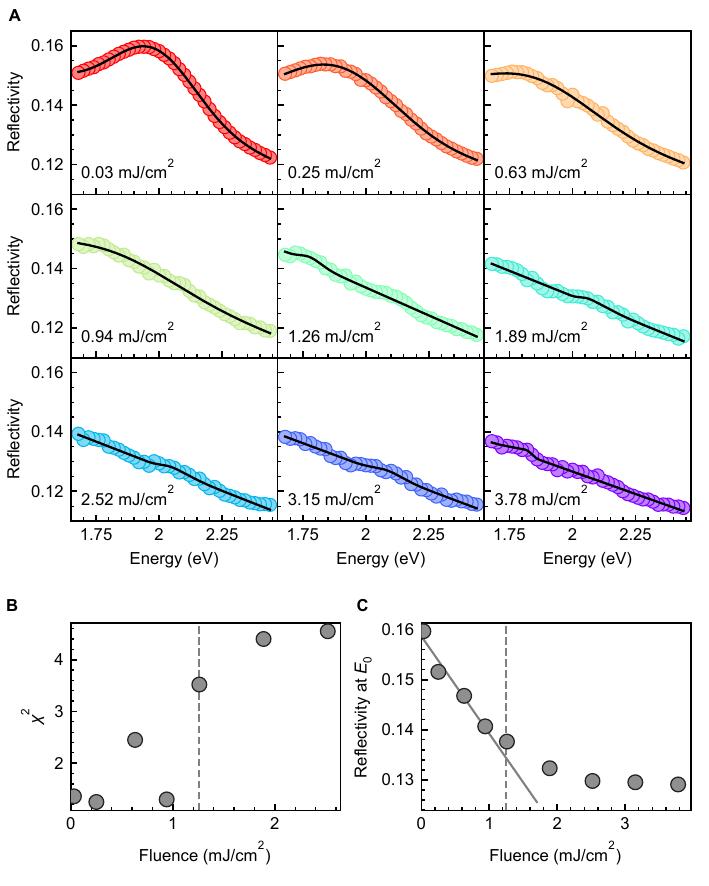}
	\caption{\textbf{Fitting of the transient reflectivity spectra near the CT transition under 400~nm pump.}(\textbf{A}) Transient reflectivity under various pump fluences (circles) overlaid with single-peak Gaussian fits (black lines). The data is the same as that in Fig.~\ref{fig:m3}A. (\textbf{B}) Reduced chi-square $\chi_\nu^2$ of the single-peak fit as a function of pump fluence. (\textbf{C}) Absolute reflectance at the equilibrium CT peak energy as a function of pump fluence. Vertical dashed lines in (\textbf{B}) and (\textbf{C}) denote the critical fluence $F^*$.}
	\label{sfig:r fitting}
\end{figure*}

We fit the reflectivity spectra near the CT transition by the sum of a linear background with a skew Gaussian function to precisely determine $F^*$, following the procedure from \cite{huang2025ultrafast}. In Fig.~\ref{sfig:r fitting}A, we show the raw data along with the fitting function. At low fluences, a single Gaussian peak accurately captures the lineshape. However, upon crossing the critical fluence $F^*$, a distinct shoulder-like feature emerges on the high energy side of the spectra, necessitating an additional Gaussian peak to achieve a satisfactory fit. 

The necessity of this two-peak model is statistically validated by the reduced chi-square $\chi_\nu^2$ metric. With the number of data points $n_d$ and free parameters $n_k$ in the model, the standard deviation of $\chi_\nu^2=\sqrt{2/(n_d-n_k)}$ is around 0.24, so a good fit scatters $1\pm0.24$. If $\chi_\nu^2$ is far larger than this, it means that the model misfits the data and uncertainties are underestimated, and if it is far smaller than this, the model overfits and uncertainties are overestimated. Figure~\ref{sfig:r fitting}B displays $\chi_\nu^2$ as a function of the pump fluence. $\chi_\nu^2$ below $F^*$ is generally much closer to 1 and smaller than those above $F^*$, thus demonstrating that a single peak misfits the data and fails to capture the spectral bifurcation.

Aside from analysis of the fitting procedure, we also present a model-free observable that confirms the presence of $F^*$. To this end, we show the absolute reflectivity at the peak energy of the CT peak as a function of pump fluence in Fig.~\ref{sfig:r fitting}C. In the weak excitation regime, the photo-response is perturbative: each absorbed pump photon excites a fixed number of carriers, and each carrier removes a fixed bit of CT oscillator strength. The bleaching of $R(E_0)$ is therefore proportional to the number of absorbed photons and linear in fluence. This holds as long as the excitations are dilute and independent, where the material responds to the pump as a simple superposition of single-photon events. Nevertheless, once the fluence surpasses $F^*$, the mobility of carriers sufficiently screens the laser field, causing the reflectance $R(E_0)$ to abruptly deviate from its low-fluence linear extrapolation. See Sec.~\ref{sec:generalizability} for detailed discussions. This confirms $F^*$ as the threshold for a nonthermal many-body electronic reconstruction.

\clearpage

\section{Absence of absorption saturation for photo-excited responses}
\label{sec:absorption saturation}
\begin{figure}[htb]
	\centering
    \includegraphics{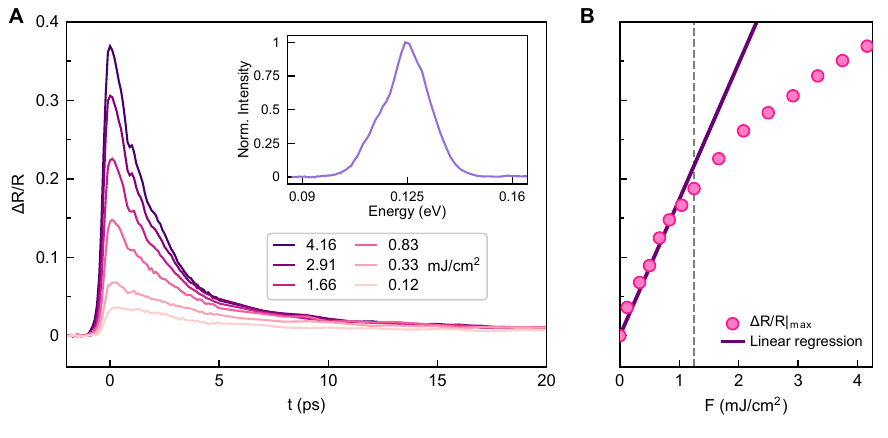}
	\caption{\textbf{Mid-IR differential transient reflectivity from La$_2$CuO$_4$ under 400~nm pump.}
	(\textbf{A}) Representative differential reflectivity transients $\Delta R/R$ probed at 0.124~eV. The inset shows the spectrum of the mid-IR probe pulse. (\textbf{B}) Pump fluence dependence of the peak $\Delta R/R$ amplitude near zero time delay. The vertical dashed line marks the critical fluence $F^*$ and the purple solid line represents linear regression of the data at low fluence.}
	\label{sfig:mir tr r}
\end{figure}

To rule out the optical absorption saturation as the origin for photo-excited HHG responses, we performed complementary transient reflectivity measurements using identical pump conditions. A 0.124~eV mid-IR pulse (Fig.~\ref{sfig:mir tr r}A, inset) was utilized to probe the low-energy dynamics following 400~nm pump excitation. Because this low probe energy is far from mid-gap absorption bands and the CT transition, the response is intrinsically isolated from effects related to band shifting or formation \cite{okamoto2011photoinduced}. 

Figure~\ref{sfig:mir tr r}A shows representative differential reflectivity transients $\Delta R/R$ at selected pump fluences. The reflectivity exhibits an increase upon pumping, followed by a picosecond decay back to equilibrium, consistent with previous literature \cite{okamoto2011photoinduced}. Notably, when tracking the peak $\Delta R/R$ signal against pump fluence (Fig.~\ref{sfig:mir tr r}B), the response continues to rise monotonically up to at least 4.16~mJ/cm$^2$, albeit deviating from a linear fashion around $F^*$ and exhibiting a slower increasing trend. This result indicates that the saturation observed in the HHG yield (Fig.~\ref{fig:m3}B, top three rows) is a distinct electronic phenomenon, rather than a trivial consequence of pump absorption saturation.

\clearpage

\section{Excitation-induced dephasing in HHG}
\label{sec:eid}
\subsection{A phenomenological model for photo-excited responses}
\label{sec:saturation model}
We discuss the role of excitation-induced dephasing (EID) and present a phenomenological model for describing nonequilibrium HHG responses in detail here. In order to rigorously evaluate possible effects induced by the above-gap pump, we first estimate the effective photo-doping level. The density of photo-excited carriers $n_\mathrm{ex}$ created by the pump pulse can be calculated by
\begin{equation}
    n_\mathrm{ex}=\frac{F(1-R)V_\mathrm{uc}}{N_\mathrm{Cu}E_\mathrm{ph}\delta}
\end{equation}
where $F$ is the pump fluence, $R=0.144$ and $\delta=61.5$~nm are the reflectivity and penetration depth at the pump wavelength $\lambda=400$~nm determined from ellipsometry measurements (Appendix~\ref{sec:reflectivity}), $V_\mathrm{uc}$ is the unit cell volume of La$_2$CuO$_4$, $N_\mathrm{Cu}$ is the number of Cu atoms per unit cell, and $E_\mathrm{ph}$ is the pump photon energy (3.1~eV). Based on this relationship, a pump fluence of 1~mJ/cm$^2$ is converted to the injection of 0.0268 electron per Cu atom.

As already discussed in the main text, injecting incoherent photo-carriers introduces three primary effects: it increases the carrier density (enhancing intraband HHG), amplifies the electron scattering rate, and triggers many-body band renormalization (in strongly correlated electron systems). We note that in literature, it has been shown that these photocarriers could also block the phase space and deplete the ground state to suppress the harmonic generation efficiency \cite{wang2017roles,cheng2020ultrafast,wang2022optical}. While the state blocking effect is always at play, we believe it has minor impact on our results because (i) the HHG susceptibility to an above-gap pump ($dI_\mathrm{HHG}/dn_\mathrm{ex}$) remains nonzero even at very low fluences, in contrast to standard semiconductor Bloch equation simulations in which it only manifests after $n$ reaches a substantial concentration \cite{wang2017roles}; in fact, in our experiment, certain harmonic orders can be completely suppressed even at a minimal photo-doping level of $\sim$1\% (Fig.~\ref{sfig:harmonic order dependence}); (ii) the 3.1~eV photon energy of our optical pump is much larger than the CT gap size; near zero time delay, these injected carriers have not yet thermalized to fully populate the conduction band minimum and valence band maximum, where the tunneling probability is the highest, and therefore should have negligible role in blocking these states.

\begin{figure}
	\centering
	\includegraphics[width=\textwidth]{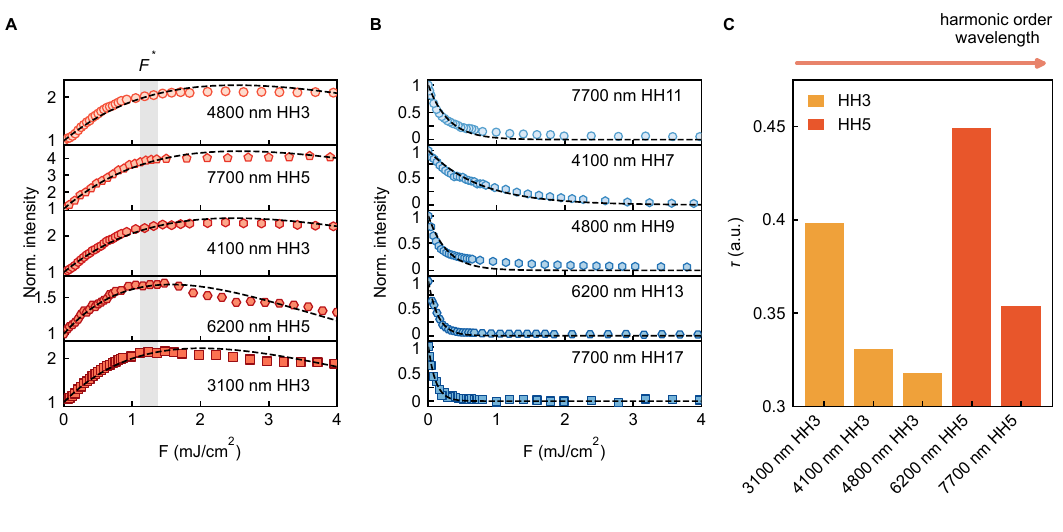}
	\caption{\textbf{Phenomenological model fit to HHG from La$_2$CuO$_4$ under 400~nm pump.}
	(\textbf{A}) Pump fluence dependence of transient HHs below the gap. (\textbf{B}) Pump fluence dependence of transient HHs above the gap. The black dashed lines in (\textbf{A}) and (\textbf{B}) denote the fit to the phenomenological model. (\textbf{C}) Extracted normalization parameters $\tau_3$ (yellow) and $\tau_5$ (orange) under varying driving conditions.}
	\label{sfig:saturation model}
\end{figure}

We then present a phenomenological model that can largely explain the nonequilibrium above-gap HHG responses while fundamentally failing to describe the below-gap orders. Such a model has been deployed for transient HHG signals in monolayer MoS$_2$ \cite{heide2022probing} and niobium oxide \cite{nie2023following}. Phenomenologically, the intensity of each interband HH decays exponentially with the dephasing rate \cite{heide2022probing,nagai2023effect}
\begin{equation}
    I_N(n)\propto\exp{(-\frac{\tau_N}{T_2(n)})}, \label{eqn:interband saturation}
\end{equation}
where $\tau_N$ is a normalization parameter for the $N$-th order and $T_2(n)$ is the carrier dephasing time constant. This exponential dependence is further verified by our numerical simulations in Sec.~\ref{sec:sbe} (Fig.~\ref{sfig:sbe dephasing dependence}A). The ratio between the harmonic intensity with and without photo-doping is expressed as
\begin{equation}
    \frac{I_N(n_\mathrm{IR}+n_\mathrm{UV})}{I_N(n_\mathrm{IR})}=\exp\big(-\tau_N(\frac{1}{T_2(n_\mathrm{IR}+n_\mathrm{UV})}-\frac{1}{T_2(n_\mathrm{IR})})\big),
\end{equation}
where $n_\mathrm{IR}$ and $n_\mathrm{UV}$ represent the carrier densities generated by the IR and UV pulses, respectively. As shown in Fig.~\ref{sfig:above gap hhg field dep}, in our experimental setting, the fluence dependence of above-gap transient HHG is independent of the applied IR field strength, i.e., $n_\mathrm{IR}$. This behavior is readily solved by assuming a linear dependence of $T_2^{-1}(n)$ on $n$. Using this scaling, the phenomenological model fits the experimental above-gap harmonic data reasonably well (black dashed lines in Fig.~\ref{sfig:saturation model}B).

On the other hand, we also tried to fit the below-gap responses using the same framework, despite the considerable pump-probe penetration depth mismatch (see discussions in Sec.~\ref{sec:giant intraband HHG}). Because the pump directly creates doublons and holons that drive intraband HHG, Eq.~\ref{eqn:interband saturation} needs to be modified to account for the increased carrier population
\begin{equation}
    I_N(n)\propto n\exp{(-\frac{\tau_N}{\tau_p(n)})}, \label{eqn:intraband saturation}
\end{equation}
where $\tau_p(n)$ is the carrier momentum relaxation time. This modified phenomenology is also supported by our numerical simulations in Sec.~\ref{sec:sbe} (Fig.~\ref{sfig:sbe dephasing dependence}B). As shown in Fig.~\ref{sfig:saturation model}A, the alignment of the experimental data with the fitting is less satisfactory, and deviation is severe for $F>F^*=1.25$~mJ/cm$^2$. More importantly, the extracted normalization parameter $\tau_N$ exhibits a physically contradictory trend. $\tau_N$ should theoretically increase with the harmonic order at the same wavelength or increase with wavelength of the same harmonic order, as the extended electron excursion time makes the trajectory more susceptible to dephasing events. Instead, the fitted $\tau_3$ decreases as the driving wavelength increases from 3100~nm to 4800~nm, and $\tau_5$ similarly decreases from 6200~nm to 7700~nm (Fig.~\ref{sfig:saturation model}C). Such behaviors indicate the breakdown of the phenomenological model for below-gap HHG, necessitating additional nonthermal effects governed by many-body effects beyond simple carrier injection and excitation-induced dephasing.

\subsection{Semiconductor Bloch equations simulations}
\label{sec:sbe}
To rigorously investigate how a photo-injected carrier population modifies high-harmonic generation in the Mott insulating state on a single-particle level, we performed optical-pump--HHG-probe simulations based on the semiconductor Bloch equations (SBEs) for an effective two-band model of the one-dimensional Hubbard system \cite{haug2009quantum,kira2011semiconductor}. The two bands represent the lower and upper Hubbard bands, with the momentum-dependent charge gap $\varepsilon_g(k)$ taken from the Bethe-ansatz excitation spectrum (Fig.~\ref{sfig:tr hhg spectrum}A, see details in Sec.~\ref{sec:subcycle analysis}) and a constant interband dipole $d_0$. In the Houston frame comoving with the vector potential, $k(t) = q + A(t)$, the microscopic interband coherence $p_q$ and pair occupation $n_q$ evolve according to
\begin{align}
\dot p_q &= -\left[\frac{1}{T_2(t)} + \frac{i}{\hbar}\,\varepsilon_g(k)\right] p_q
            - \frac{i}{\hbar}\, d_0 E(t) \left(1 - 2 n_q\right), \\
\dot n_q &= -\frac{2}{\hbar}\,\mathrm{Im}\!\left[ d_0 E(t)\, p_q^{*} \right]
            - \frac{n_q - \bar n_q(t)}{\tau_p(t)},
\end{align}
where $T_2$ and $\tau_p$ are pair dephasing and momentum relaxation time, respectively. The emitted optical field is extracted from the total current $j = j_{\mathrm{inter}} + j_{\mathrm{intra}}$, defined as 
\begin{equation}
    j_{\mathrm{inter}} = \partial_t \!\int\! \frac{dq}{2\pi}\, 2\,\mathrm{Re}[d_0^* p_q],\quad j_{\mathrm{intra}} = \int\! \frac{dq}{2\pi}\, \Delta v(k)\, n_q,
\end{equation}
where $\Delta v = \partial_k \varepsilon_g/\hbar$ is the doublon-holon relative velocity. Harmonic spectra are then computed from the current radiated during the IR-pulse window.

Dissipation is incorporated into the SBEs through two distinct density-dependent channels. First, the interband coherence decays with an EID rate $1/T_2(t) = 1/T_2^{(0)} + \alpha n(t)$, where $n(t)$ is the instantaneous pair density per site and $\hbar\alpha$ parameterizes the carrier-density-dependent scattering. Second, momentum relaxation drives the population $n_q$ towards a thermal (Boltzmann) distribution $\bar n_q(t)$ evaluated at the instantaneous crystal momentum $k = q + A(t)$, whose normalization and effective temperature are chosen at every instant to conserve both the total pair number and the kinetic energy of the distribution. This term therefore relaxes only the shape and field-induced drift of the carrier distribution, which is exactly a Drude damping of the intraband current at rate $1/\tau_p$, without creating, destroying, or artificially cooling carriers. This relaxation rate features the same density dependence as the dephasing, $1/\tau_p(t) = 1/\tau_p^{(0)} + \beta n(t)$, where $\hbar\beta$ accounts for doublon-holon drag and umklapp scattering, which unlike normal carrier-carrier collisions degrade the charge current. Throughout the simulations, parameters are strictly constrained by the physical consistency bound $1/T_2 \geq 1/2\tau_p$ (since momentum scattering inherently causes dephasing, whereas pure dephasing does not necessarily relax momentum).

\begin{figure}[htb]
	\centering
	\includegraphics{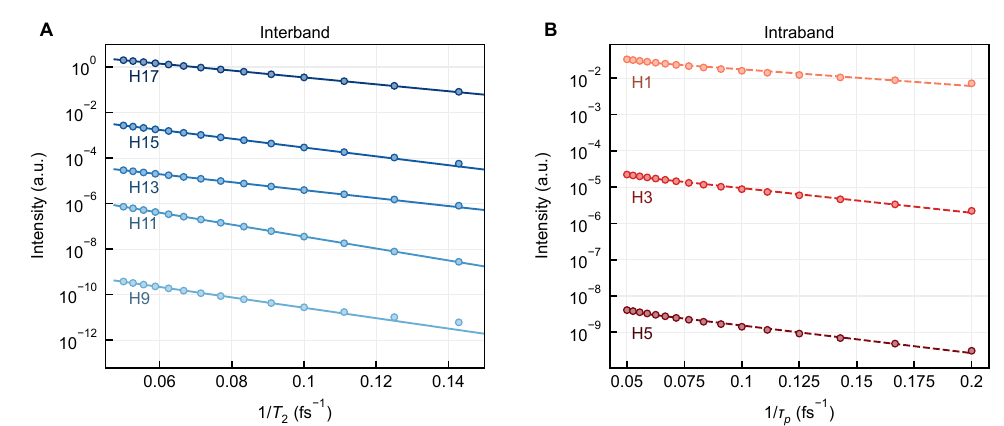}
	\caption{\textbf{Dependence of the interband and intraband harmonic yields on the dephasing and intraband relaxation rates.}
	(\textbf{A}) Integrated interband harmonic intensity as a function of the dephasing rate $1/T_2$ for harmonics HH9–HH17. Successive harmonics are offset vertically for clarity. Solid lines are single-exponential fits to the slow-dephasing data ($T_2 \ge 10$~fs). $1/\tau_p$, $\alpha$, and $\beta$ are set to zero in this set of simulations. (\textbf{B}) Integrated intraband harmonic intensity as a function of the intraband relaxation rate $1/\tau_p$ for HH1–HH5, at fixed $T_2 = 10$~fs. Dashed lines are single-exponential fits. $\alpha$ and $\beta$ are set to zero in this set of simulations.}
	\label{sfig:sbe dephasing dependence}
\end{figure}

Before introducing the optical pump, we first analyzed the isolated roles of $T_2$ and $\tau_p$ by computing the decoupled interband and intraband HHG spectra. By scanning the dephasing time $T_2$ under a fixed IR drive (0.16~eV photon energy, $E_0=0.005$~a.u., 100~fs intensity FWHM), we confirm that the interband harmonics are exponentially dependent on the dephasing rate as shown in Fig.~\ref{sfig:sbe dephasing dependence}A. Likewise, under the same drive conditions, momentum relaxation drives an exponential suppression of the intraband harmonic intensity (Fig.~\ref{sfig:sbe dephasing dependence}B). Together, these numerical results validate the phenomenological scaling relationships utilized in Eqs.~\eqref{eqn:interband saturation} and \eqref{eqn:intraband saturation}.

\begin{figure}[hb]
	\centering
	\includegraphics{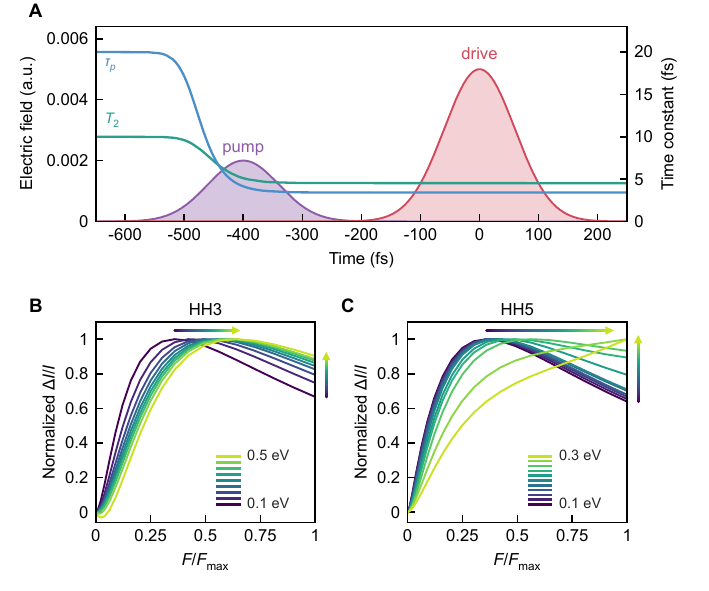}
	\caption{\textbf{Photodoping-induced modification of the intraband harmonic yield.}
	(\textbf{A}) Pulse sequence and the resulting relaxation dynamics. (\textbf{B}), (\textbf{C}) Normalized intraband harmonic enhancement $\Delta I/I$ as a function of normalized fluence $F/F_\mathrm{max}$. In (\textbf{B}), we used $E_\mathrm{IR} = 1\times10^{-3}$ a.u. (nine IR photon energies, 0.1–0.5~eV, evenly spaced); in (\textbf{C}), we used $E_\mathrm{IR} = 2.5\times10^{-3}$ a.u. (eleven IR photon energies, 0.1–0.3~eV, evenly spaced).}
	\label{sfig:sbe photodoping}
\end{figure}

Next, we simulated the HHG spectra after photo-carrier creation by an above-gap pump pulse. A 2~eV pump (100~fs intensity FWHM), arriving before the IR drive, injects carriers that shorten both dephasing and momentum relaxation times through the density-dependent scattering pathways built into the model (Fig.~\ref{sfig:sbe photodoping}A; $\hbar\alpha=0.35$~eV, $\hbar\beta=0.7$~eV). At the highest simulated pump fluence, the dephasing time falls from $T_2=10$~fs to 4.5~fs and the momentum relaxation time from $\tau_p=20$~fs to 3.4~fs. The choices of these initial values align with typical numbers in literature \cite{vampa2014theoretical,nagai2023effect}.

The resulting intraband response is highly non-monotonic. Figure~\ref{sfig:sbe photodoping}B tracks the normalized enhancement $\Delta I/I$ of HH3 across nine drive photon energies between 0.1 and 0.5~eV. The harmonic yield initially grows steeply with fluence, reaches a maximum, and subsequently declines. This turnover originates from the competition between the injected carrier density, which increases the intraband current, and the suppressed $\tau_p$, which exponentially kills it (Fig.~\ref{sfig:sbe dephasing dependence}B). Crucially, however, the critical fluence at which the yield peaks shifts systematically to higher fluence with increasing drive photon energy, from $F/F_\mathrm{max}=0.36$ at 0.1~eV to 0.64 at 0.5~eV (horizontal arrow in Fig.~\ref{sfig:sbe photodoping}B), while the enhancement retained at the highest fluence grows correspondingly (vertical arrow in Fig.~\ref{sfig:sbe photodoping}B). This drive-dependent trend becomes even more pronounced for HH5 (Fig.~\ref{sfig:sbe photodoping}C), where the maximum peak moves from $F/F_\mathrm{max}=0.36$ at 0.1~eV to beyond the scanned fluence range at higher drive energies. Because the suppression mechanism in the SBEs model is governed by the ratio of the relaxation to the drive period, the balance point between carrier gain and EID suppression is inherited from the external drive conditions. By contrast, in the experimental data, the below-gap harmonics saturate and decline at a universally constant critical fluence $F^*\approx1.25$~mJ/cm$^2$ at all driving wavelengths (Fig.~\ref{fig:m3}B). Furthermore, this exact same $F^*$ marks the bifurcation of the CT resonance in transient reflectivity. We therefore conclude that EID at the single-particle level, while successfully capturing the qualitative competition between carrier injection and relaxation, fails to rationalize the saturation dynamics of the below-gap HHG. The observed wavelength-independent threshold fundamentally demands an intrinsic electronic mechanism triggered at a critical photo-carrier density, which we identify with the reorganization of the many-body manifolds (Fig.~\ref{fig:m4}).

\clearpage

\section{Harmonic order dependence of above-gap HHG suppression}
\begin{figure}[hb]
	\centering
	\includegraphics{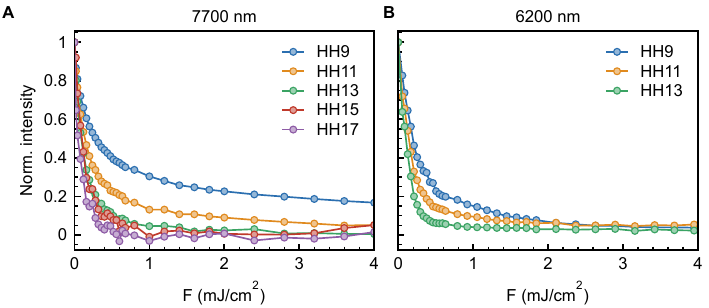}
	\caption{\textbf{Harmonic order dependence of suppressed above-gap HHG.}
	(\textbf{A}) Pump fluence dependence of transient HHs above the gap for a 7700~nm drive. (\textbf{B}) Same as (\textbf{A}) but for a 6200~nm drive.}
	\label{sfig:harmonic order dependence}
\end{figure}

Evaluating the suppression of above-gap HHG as a function of the harmonic order provides additional support for the dephasing mechanism. As detailed in Sec.~\ref{sec:saturation model}, higher order harmonics driven at a fixed wavelength should exhibit greater susceptibility to photo-doping owing to their increased sensitivity to dephasing events. This trend is directly corroborated by our fluence-dependent measurements. For harmonics generated by a 7700~nm drive, the emission undergoes progressively stronger suppression at higher orders for any given pump fluence (Fig.~\ref{sfig:harmonic order dependence}A). Similarly, Fig.~\ref{sfig:harmonic order dependence}B shows the fluence dependence for a 6200~nm drive, which is in line with the phenomenological model for the above-gap harmonics.

\clearpage

\section{Time-dependent exact diagonalization simulations}
\subsection{The Hubbard model and simulation approaches}
\label{sec:hubbard simulations}
We simulate the pulsed laser field with a time-dependent electric field (vector potential) along the $x$ direction
\begin{equation}
    A(t) = A_0 f(t)\cos{\Omega(t-t_0)}
\end{equation}
where $f(t)=e^{-(t-t_0)^2/2\sigma_t^2}$ is the Gaussian envelope function of the pulse. We match experimental conditions by setting the intensity full width at half maximum to 100~fs. Starting from a single-band 1D Fermi-Hubbard model, the pump field enters the Hamiltonian via a Peierls' substitution:
\begin{equation}
    H(t)=-t_h\sum_{j\sigma}[e^{-iA(t)}c_{j\sigma}^\dagger c_{j+1\sigma}+e^{iA(t)}c_{j+1\sigma}^\dagger c_{j\sigma}] + U\sum_{j}n_{j\uparrow}n_{j\downarrow}.
\end{equation}
Here, $c_{j\sigma}^\dagger$ ($c_{j\sigma}$) creates (annihilates) an electron at site $j$ with spin $\sigma$. The parameter $t_h$ is the nearest-neighbor hopping integral and $U$ is the on-site Coulomb repulsion. $U=8t_h$ and $t_h=0.35$~eV are set to simulate strong correlations in cuprates \cite{pavarini2001band,yoshida2006systematic}. We choose a chain size $L=12$ and a coarse-grained time set $\delta t=0.05t_h^{-1}$.

We use the parallel Arnoldi method \cite{lehoucq1998arpack} to determine the ground-state wavefunction \cite{jia2018paradeisos} and the Krylov subspace technique \cite{balzer2011krylov,park1986unitary,hochbruck1997krylov,moler2003nineteen} to evaluate the evolution of a state $|\psi(t+\delta t)\rangle=e^{-iH(t)\delta t}|\psi(t)\rangle$. We confirm the convergence of real-time propagation by comparing the current and HHG spectra with increasing Krylov subspace dimension and decreasing time steps.

To compute the harmonic emission, we first use the electric current operator defined as
\begin{equation}
    J(t) = -it_h\sum_{j\sigma}[e^{-iA(t)}c_{j\sigma}^\dagger c_{j+1\sigma}-e^{iA(t)}c_{j+1\sigma}^\dagger c_{j\sigma}]
\end{equation}
to calculate the time-dependent current. HHG spectrum from charge acceleration is then given by
\begin{equation}
    S(\omega) = |\int f(t)\partial_t \langle J(t)\rangle e^{-i\omega t}dt|^2.
\end{equation}
The simulated HHG spectra in Fig.~\ref{fig:m4}B display continuum-like spectral structures near the plateau emission energy, which may arise from (1) finite duration of the pulse that breaks strict time periodicity; and (2) long coherence time of the doublon-holon pair due to the absence of dephasing channels (for example, from electron-phonon coupling) that leads to interference among multiple collision events along quasiparticle trajectories. This observation is fully consistent with previous reports of HHG from a Hubbard model \cite{silva2018high,orthodoxou2021high} and has been extensively discussed elsewhere \cite{murakami2021high,murakami2022anomalous,vampa2015semiclassical}.

To characterize the evolution of elementary excitations, we calculate time-dependent single-particle spectral function of the form \cite{freericks2009theoretical}
\begin{equation}
    \mathcal{A}(k,\omega;t) = \frac{1}{2\pi i}\iint e^{i\omega(t_1-t_2)}g(t_1;t)g(t_2;t)G_k^<(t_1,t_2)dt_1dt_2
\end{equation}
where $g(t_0;t)$ describes the probe pulse envelope and is given by a Gaussian function centered at $t_0$ with standard deviation $\sigma_p$
\begin{equation}
    g(t_0;t) = \frac{1}{\sqrt{2\pi}\sigma_p}e^{-(t-t_0)^2/2\sigma_p^2}
\end{equation}
and $G_k^<(t_1,t_2)$ is the two-time nonequilibrium lesser Green's function given by
\begin{equation}
    G_k^<(t_1,t_2) = i\theta(t_2-t_1)\langle c_k^\dagger(t_2)c_k(t_1)\rangle.
\end{equation}
Due to the Fourier transform limitation and the need to highlight the dynamical properties, we choose a probe width $\sigma_p=2t_h^{-1}$ as a compromise between time and frequency resolutions.

In both 1D and 2D, the equilibrium Hubbard bands spectral function $\mathcal{A}(k,\omega)$ is dressed by coupling to the spin sector, albeit through different microscopic mechanisms:
\begin{itemize}
\setlength{\itemsep}{0pt}
\setlength{\parskip}{0pt}
\setlength{\parsep}{0pt}
\item 1D Hubbard chain ($U/t \gg 1$): the Bethe ansatz factorizes the eigenstates into independent charge (holon) and spin (spinon) sectors. The electron creation operator $c^{\dagger}_{k\sigma}$ generates a superposition of holon ($\sim t$) and spinon ($\sim J = 4t^{2}/U$) excitations, and $\mathcal{A}(k,\omega)$ is the resulting kinematic continuum bounded by the two branches (Fig.~\ref{sfig:hubbard spectral function}A, the leftmost panel).
\item 2D Hubbard model on a square lattice: the half-filled ground state hosts an AFM order with a long correlation length. A propagating doublon misaligns spins along its trajectory at a cost of $\sim J$ per hop, and is dressed into a spin polaron of bandwidth $\sim J$ accompanied by an incoherent multi-magnon shoulder.
\end{itemize}
While our major results were conducted on the 1D Hubbard chain, we argue in Sec.~\ref{sec:generalizability} that the general organizing principle does not involve physics unique to 1D Hubbard systems such as spin-charge separation.

The single-band Hubbard model with nearest-neighbor hopping on a bipartite lattice is invariant under the particle-hole transformation $c_{i\sigma} \to (-1)^{i}\,c^{\dagger}_{i\bar{\sigma}}$, which exchanges doublons and holons and maps the UHB onto the LHB. Therefore, the removal spectral function $\mathcal{A}_-(k,\omega)$ of the UHB is equivalent (up to a momentum shift) to the addition spectral function $\mathcal{A}_+(k,\omega)$ of the LHB. We thus only present $\mathcal{A}_-(k,\omega)$ for the UHB throughout Fig.~\ref{fig:m4} and Figs.~\ref{sfig:hubbard spectral function}-\ref{sfig:hubbard dos}; the corresponding holon evolution in the LHB is identical by symmetry.

\subsection{System size dependence of the microscopic model}
\label{sec:system size dependence}
Due to the demands to calculate $\mathcal{A}(k,\omega;t)$, whose time cost scales aggressively with system size, our full study is based on a 12-site chain. To ensure these results are robust against finite-size effects, we compared the charge dynamics across 12-, 14- and 16-site chains.

\begin{figure}[hb]
	\centering
	\includegraphics{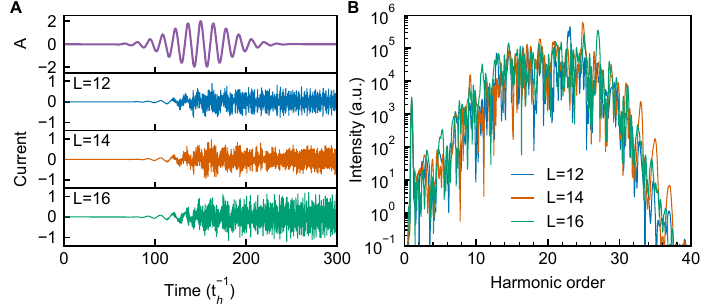}
	\caption{\textbf{Hubbard chain length dependence of HHG.}
	(\textbf{A}) Time-domain vector potential (top panel) and corresponding current (bottom panels) for varying chain lengths labeled in each panel. $A = 2$ and $\Omega = 0.45t_h$ are utilized in this simulation. (\textbf{B}) HHG spectra from the time-dependent current across different chain lengths.}
	\label{sfig:system size dependence}
\end{figure}

Figure~\ref{sfig:system size dependence}A shows the current under the same external drive in three systems. Although the overall amplitudes vary slightly, the fundamental HHG spectra exhibit identical characteristics (Fig.~\ref{sfig:system size dependence}B), namely, a strong linear response, a plateau near the Mott transition energy, and a cutoff beyond the maximal Hubbard bands separation. Therefore, finite-size effects do not alter our conclusions from the microscopic model.

\subsection{Non-perturbative nature of HHG from the Hubbard model}
\begin{figure*}[htb]
	\centering
	\includegraphics{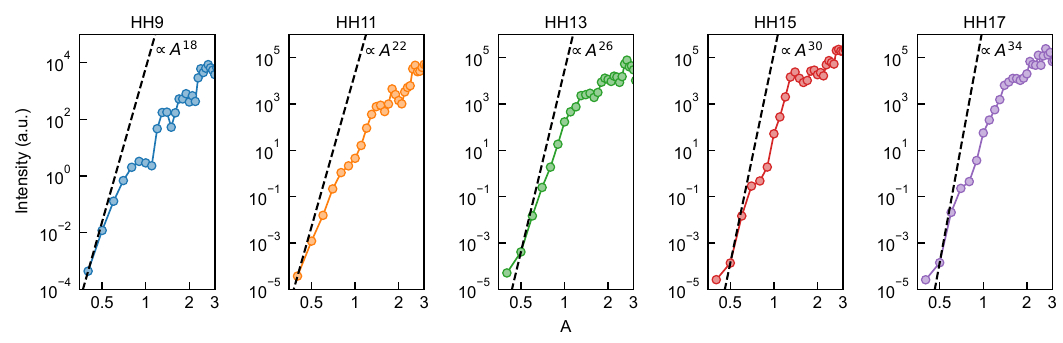}
	\caption{\textbf{Non-perturbative field scaling of HHG from a single-band Hubbard model.}
	Dependence of the harmonic intensity on the peak vector potential of $\Omega = 0.45t_h$ drive on a log-log scale. The dashed line in each panel displays the expected perturbative power-law scaling at low field strength.}
	\label{sfig:hubbard field dep}
\end{figure*}

We verify the non-perturbative character of HHG from the Hubbard model by analyzing the harmonic yield as a function of the peak driving field. Due to the lack of well-defined HH peaks at high fields (Fig.~\ref{sfig:hubbard cutoff}A), we utilize the integrated emission intensity around the odd-order harmonic frequencies with the widths of the fundamental frequency as the HHG intensity. Figure~\ref{sfig:hubbard field dep} shows the strength of HHs by $\Omega = 0.45t_h$ drive as a function of the peak vector potential on a log-log scale, revealing a clear deviation from the perturbative nonlinear optics scaling beyond the low-field regime (black dashed lines). Thus, for the field strength $A_0=2$ employed in our simulations, HHG originates from the non-perturbative nonlinear charge dynamics within the driven system.

\subsection{HHG cutoff in the single-band Hubbard model}
We discuss the consistency between the cutoff behavior observed in experiments and the single-band Hubbard model simulations here. In solid-state HHG, the cutoff is often defined as the highest order observable in experiments. We followed this convention in the present work. Nevertheless, numerical simulations usually have much higher accuracy so that very high orders much beyond the plateau could be resolved, which is also the case in our data. As a result, we define the cutoff in simulations as the order from which the harmonic intensity starts to drop exponentially, following the convention of gas-phase HHG \cite{lewenstein1994theory}. 

\begin{figure*}[htb]
	\centering
	\includegraphics{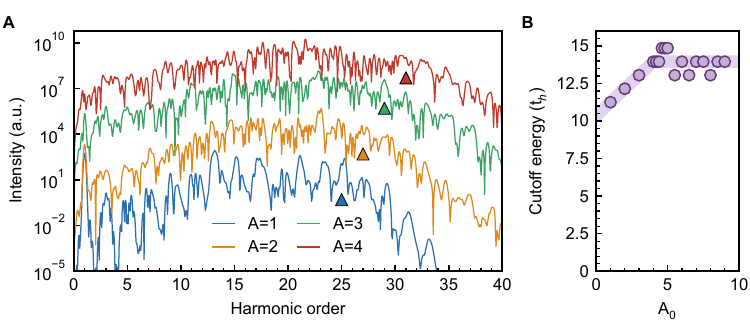}
	\caption{\textbf{Cutoff behavior of HHG from a single-band Hubbard model.}
	(\textbf{A}) Representative HHG spectra at selected peak fields with corresponding cutoff orders labeled by the triangle. $\Omega = 0.45t_h$ is adopted in this simulation to acquire a high frequency resolution of the cutoff energy. Each spectrum is vertically offset for clarity. (\textbf{B}) Peak field $A_0$ dependence of the extracted cutoff. The shaded purple line shows the linear regression of cutoff energy at low fields ($A_0\le4$) and the constant cutoff at high fields ($A_0 > 4$).}
	\label{sfig:hubbard cutoff}
\end{figure*}

Figure~\ref{sfig:hubbard cutoff}A shows representative simulated HHG spectra at different field strengths and the cutoff order in each spectrum is labeled by a triangle. In Fig.~\ref{sfig:hubbard cutoff}B, we plot the extracted cutoff order as a function of the peak field. At low field strengths, the cutoff scales linearly with the driving field, consistent with previous density matrix renormalization group studies of 1D Mott insulators \cite{murakami2021high}. However, in the strong-field regime, when the cutoff reaches the maximal separation of lower and upper Hubbard bands, it abruptly saturates and stops growing. This behavior is consistent with experimental observations and supports the interband nature of these HHs above the Mott gap.

\subsection{Time-frequency spectrogram analysis}
\label{sec:subcycle analysis}
We analyze the time-domain HHG dynamics to elucidate the interband and intraband nature of above-gap and below-gap HHs in the driven 1D Hubbard model, respectively. To gain insights into this, we perform a subcycle decomposition of the current $J(t)$ by wavelet analysis to extract the temporal emission amplitude at each frequency. Time-domain HHG is computed by continuous wavelet transform
\begin{equation}
    I(s,\tau)=\frac{1}{\sqrt{|s|}}\int_{-\infty}^{\infty}I(t)\psi^*(\frac{t-\tau}{s})dt
\end{equation}
where $I(t)=\partial_t \langle J(t)\rangle$ is the charge acceleration used to calculate the HHG spectrum and $\psi$ is the wavelet. We use a Morlet wavelet to perform the analysis
\begin{equation}
    \psi(u)=\pi^{-\frac{1}{4}}e^{i\omega_0u}e^{-\frac{u^2}{2}}
\end{equation}
where $\omega_0=7.5$ (the dimensionless Morlet central frequency) is adopted to ensure a simultaneous high enough time and frequency resolution.

\begin{figure*}[htb]
	\centering
	\includegraphics{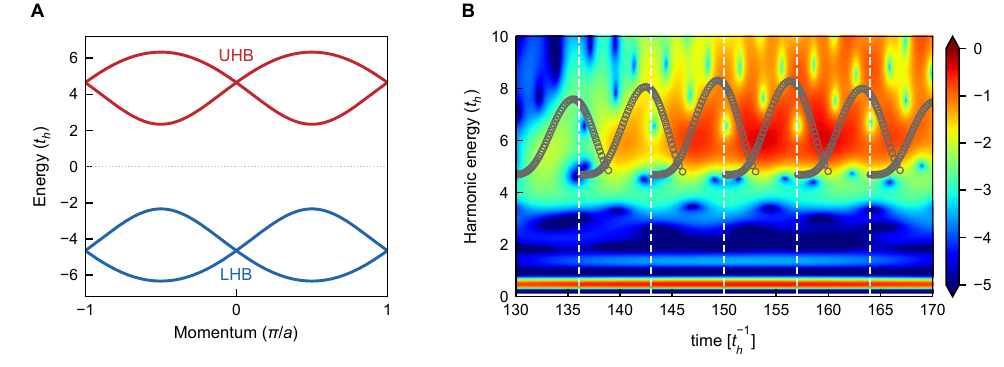}
	\caption{\textbf{Time-domain HH emission.}
	(\textbf{A}) Holon dispersion of the half-filled 1D Hubbard model utilized in the semiclassical analysis. (\textbf{B}) Spectrogram of HHG on a log scale near the center of the pulse. The grey circles show the recombination times from semiclassical analysis. The white dashed lines denote the times when $E=0$. $\Omega = 0.45t_h$ and $A=1$ are adopted in this simulation.}
	\label{sfig:tr hhg spectrum}
\end{figure*}

The semiclassical model equations are already listed in Eqs.~\eqref{si:semi1} and \eqref{si:semi2} in Sec.~\ref{sec:finite excursion length}. The holon dispersion of 1D Hubbard model at half filling can be analytically written by parameterization through rapidity $k$, and consequently the momentum $p_{h,d}(k)$ and energy $\varepsilon_{h,d}(k)$ are given by
\begin{align}
    p_h(k)&=p_d(k)+\pi=\frac{\pi}{2}-k-2\int_0^\infty\frac{\mathcal{J}_0(\omega)\sin{(\omega\sin{k})}}{1+e^{\frac{\omega U}{2}}}\frac{d\omega}{\omega}\label{eq:1d hubbard p},\\
    \varepsilon_{h}(k)&=\varepsilon_{d}(k)=\frac{U}{2}+2\cos{k}+2\int_0^\infty\frac{\mathcal{J}_1(\omega)\cos{(\omega\sin{k})}e^{-\frac{\omega U}{4}}}{\cosh{\frac{\omega U}{4}}}\frac{d\omega}{\omega}\label{eq:1d hubbard e},
\end{align}
where $\mathcal{J}_n$ is the $n$-th Bessel function of the first kind \cite{essler2005one}. With $U/t_h=8$, the holon dispersion is shown in Fig.~\ref{sfig:tr hhg spectrum}A.

Figure~\ref{sfig:tr hhg spectrum}B shows the calculated recombination time and energy (grey markers) overlaid on top of the HHG spectrogram. Strong qualitative agreement is yielded, although the semiclassical analysis tends to slightly overestimate (underestimate) the light frequency in the first (second) half of an optical cycle. This chirp indicates that the HHs in the above-gap regime originate from the recombination of doublons and holons. Conversely, the low-order HHs (1st and 3rd order in the present case) barely exhibit any chirping and different orders are generated simultaneously, confirming their intraband nature from dynamical Bloch oscillations \cite{wu2015high}. Taken together, we conclude that the present subcycle analysis delineates the essential features of nonlinear doublon-holon dynamics and their contribution to HHG, in full consistency with a previous study of 1D Mott insulators based on exact diagonalization and the infinite time-evolving block decimation method \cite{murakami2021high}.

\subsection{Doublon-holon number dynamics during HHG}
\begin{figure}[htb]
	\centering
	\includegraphics{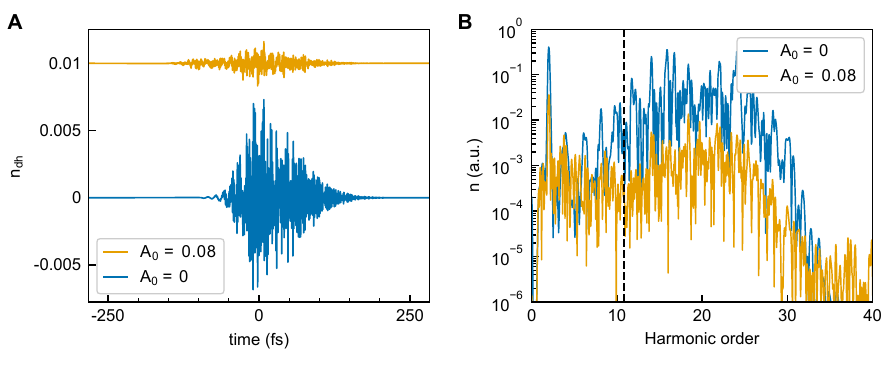}
	\caption{\textbf{Real-time doublon-holon number dynamics.}
    (\textbf{A}) ac component of the carrier number dynamics before (blue) and after (orange) the pump. The two curves are vertically offset for clarity. (\textbf{B}) Power spectral density of the carrier dynamics in (\textbf{A}). The black dashed line denotes the Mott gap energy. $\Omega = 0.45t_h$, $A=2$, and $\omega_\mathrm{pump} = 8.86t_h$ are adopted in this simulation.}
	\label{sfig:hubbard carrier dynamics}
\end{figure}

To corroborate the assignment of HHG channels from subcycle analysis in Sec.~\ref{sec:subcycle analysis}, we tracked the doublon-holon number dynamics during the HHG process. Figure~\ref{sfig:hubbard carrier dynamics}A shows the real-time evolution of doublon-holon pairs during the drive, after filtering out the dc component that corresponds to a bare increase in their number and multiplying the pulse envelope. The addition of a pump pulse ($\Omega_\mathrm{pump} = 8.86t_h$, $A_0=0.08$) severely dampens nonlinear oscillations in the charge sector. Their power spectral density (Fig.~\ref{sfig:hubbard carrier dynamics}B) reveals an overall amplitude decrease of the charge dynamics across all sub-cutoff frequencies. The responses in the carrier number suggest that the interband HHG channel, which comes from coherent creation and annihilation of doublons and holons, is generally suppressed upon photo-doping. Thus, the observed enhancement for below-gap harmonics in Fig.~\ref{fig:m4} must arise from an intraband process, which inherently preserves the number of carriers and therefore cannot be directly resolved here.

\subsection{Driving electric field dependence of transient HHG in the Hubbard model}
\begin{figure}[htb]
	\centering
	\includegraphics{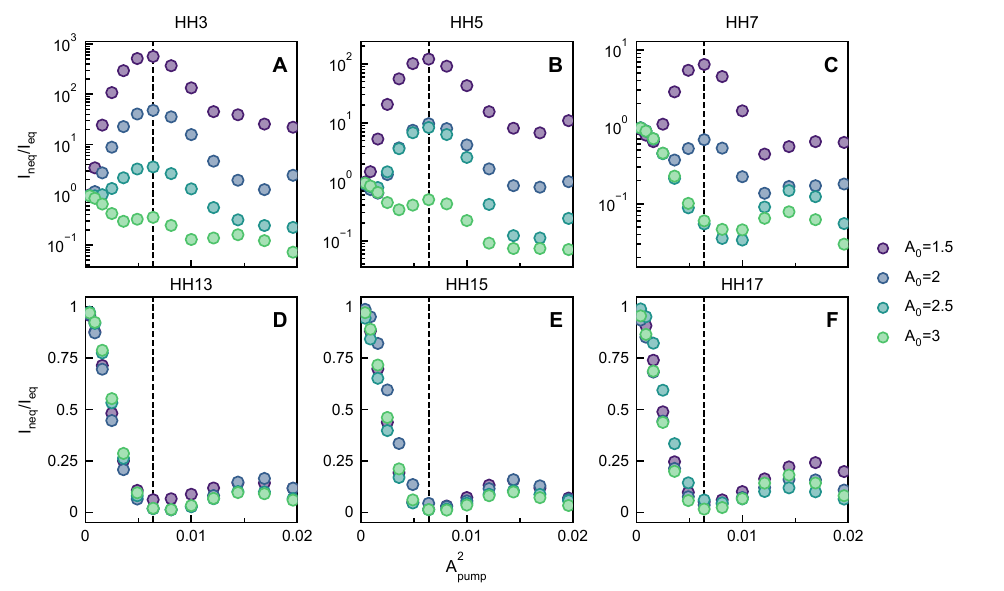}
	\caption{\textbf{Driving-field dependence of the transient harmonic response.}
    (\textbf{A}-\textbf{C}) Normalized intensity of below-gap harmonics as a function of pump fluence for varying drive field strengths. The $y$-axis is on a logarithmic scale. (\textbf{D}-\textbf{F}) Normalized intensity of above-gap harmonics as a function of pump fluence for varying drive field strengths. $\Omega = 0.45t_h$ and $\omega_\mathrm{pump} = 8.86t_h$ are adopted in this simulation.}
	\label{sfig:hubbard noneq drive field dep}
\end{figure}

We show the dependence of transient HHG responses on the peak driving electric field here. Figure~\ref{sfig:hubbard noneq drive field dep}A--C show the fluence dependence of below-gap HHG at various drive electric field strengths, and it can be clearly seen that the enhancement behavior is more prominent at lower peak field, consistent with the experimental trends (Fig.~\ref{sfig:hh5 enhancement}). In contrast, the above-gap harmonics are largely insensitive to driving fields (Fig.~\ref{sfig:hubbard noneq drive field dep}D to F), again echoing the experimental trends (Fig.~\ref{sfig:above gap hhg field dep}) as well as the phenomenological model presented in Sec.~\ref{sec:saturation model}.

\subsection{Evolution of single-particle spectral function across the light-induced band renormalization transition}
\begin{figure}[h]
	\centering
	\includegraphics[width=0.85\textwidth]{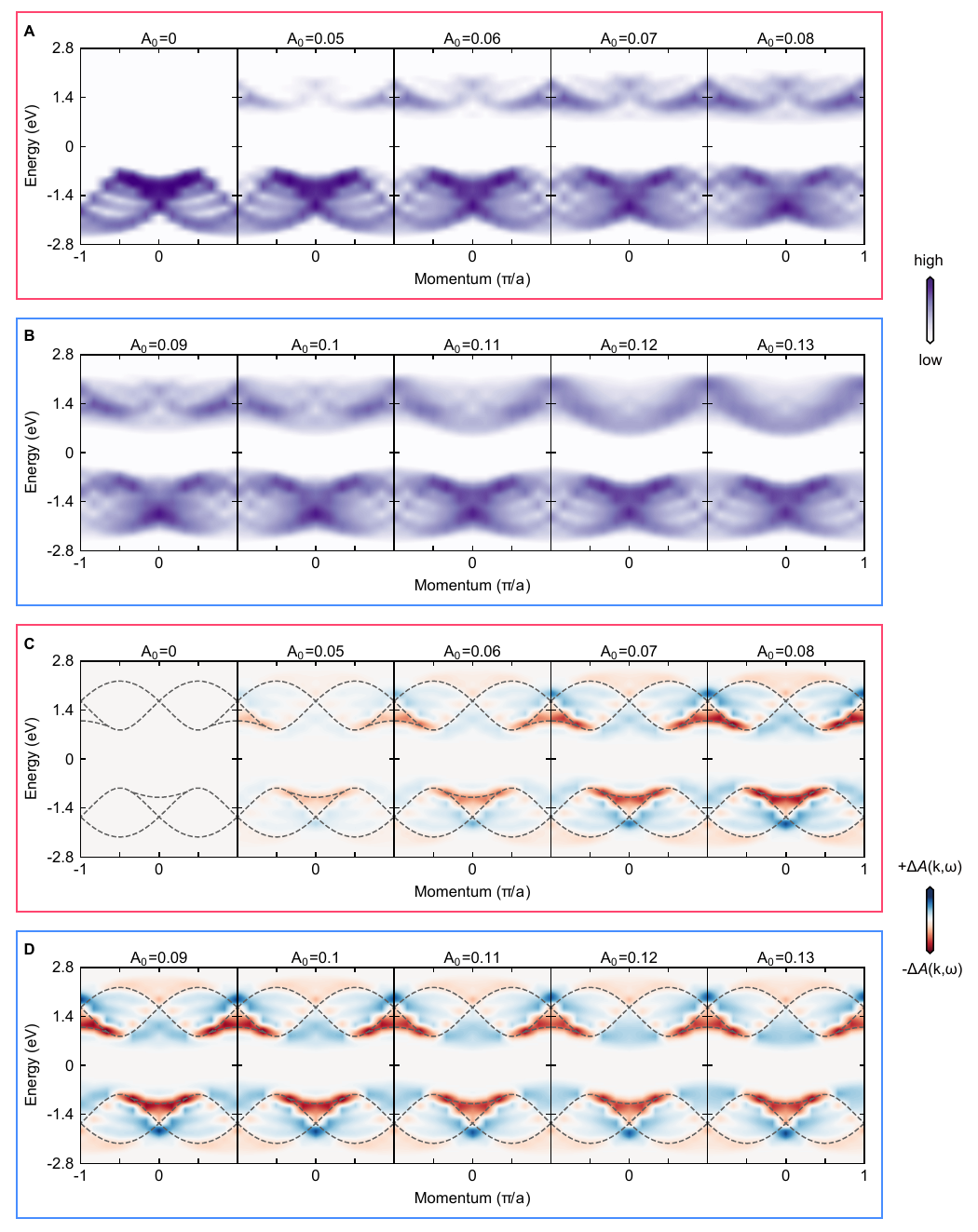}
	\caption{\textbf{Evolution of single-particle spectral functions across the light-induced band reconstruction transition.}
    (\textbf{A}-\textbf{B}) Single-particle removal spectral function as a function of the pump peak vector potential below (\textbf{A}) and above (\textbf{B}) the threshold. The colormap is on a log scale. (\textbf{C}-\textbf{D}) Difference between pumped and static total spectral function as a function of the pump peak vector potential below (\textbf{C}) and above (\textbf{D}) the threshold. The colormap is on a linear scale.}
	\label{sfig:hubbard spectral function}
\end{figure}

We show the detailed evolution of single-particle spectral functions across the light-induced band renormalization transition in Fig.~\ref{sfig:hubbard spectral function}, complementary to the results in Fig.~\ref{fig:m4}F-H. When the pump strength is lower than the threshold ($A_0<A_{0c}$), UHB is sparsely populated (Fig.~\ref{sfig:hubbard spectral function}A) and the band dispersion change is negligible (Fig.~\ref{sfig:hubbard spectral function}C). At the critical field ($A_0=A_{0c}$), finite spectral weight begins to accumulate near the $\Gamma$ point beyond the holon branch, as seen in both removal spectral function $\mathcal{A}^-$ and the difference in total spectral function $\Delta\mathcal{A}$ (Fig.~\ref{sfig:hubbard spectral function}, A and C). Lastly, exceeding the threshold ($A_0>A_{0c}$), UHB fundamentally restructures into a parabolic-like shape (Fig.~\ref{sfig:hubbard spectral function}, B and D), corresponding directly to the non-monotonic HHG yield.

\begin{figure}[htb]
	\centering
	\includegraphics{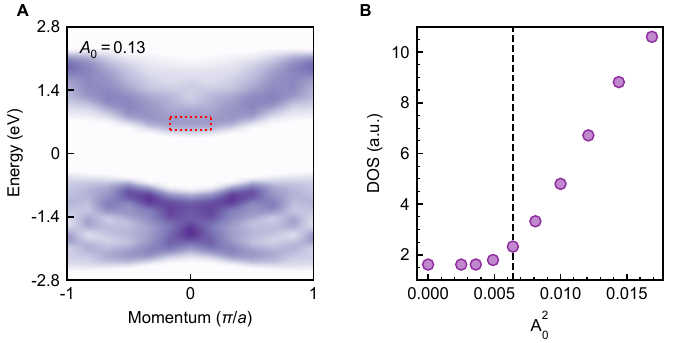}
	\caption{\textbf{Single-particle spectral weight evolution near the $\Gamma$ point.}
    (\textbf{A}) Single-particle removal spectral function $\mathcal{A}^-$ at $A_0=0.13$. The red box shows the region for integrating spectral weight in (\textbf{B}). (\textbf{B}) Density of states (DOS) of the reconstructed band near the $\Gamma$ point as a function of pump fluence. The black dashed line marks the critical threshold $A_{0c}^2$.}
	\label{sfig:hubbard dos}
\end{figure}

To analyze this process quantitatively, we integrate the spectral weight near the renormalized UHB bottom outside the equilibrium dispersion range (red box in Fig.~\ref{sfig:hubbard dos}A). The density of states within this momentum-energy space as a function of pump fluence shows a clear threshold at $A_\mathrm{0c}^2$ (Fig.~\ref{sfig:hubbard dos}B), corroborating the transition observed in time-resolved HHG.

\subsection{Generalizability of the microscopic model to La$_2$CuO$_4$}\label{sec:generalizability}
The numerical simulations in the main text are performed on a half-filled 1D Hubbard chain, while our experiment probes the quasi-2D cuprate Mott insulator La$_2$CuO$_4$. Here we discuss the universal physical mechanism invoked in the main text and explain why it carries over from 1D to 2D despite microscopically distinct equilibrium dispersions. We further suggest that the difference in crossover sharpness between simulation and experiments is qualitatively consistent with a simple geometric counting.

The intraband HHG channel conserves the number of excited carriers and therefore corresponds to doublons and/or holons hopping to their neighboring singly occupied site in real space. At low fluence, the photogenerated doublons and holons are spatially correlated as nearest-neighbor pairs, because the dipole operator in our model only couples nearest-neighbor sites through the Peierls phase on the hopping; once created, a residual spin-mediated attraction keeps them adjacent against drift. Above $A_{0c}$, this constraint breaks down — the spin-disordered region around each pair overlaps with that of its neighbors, and pair-breaking events become statistically dominant. The doublons and holons then propagate independently across the lattice. Since they are not adjacent to each other, intraband hopping takes place along any direction, effectively transferring the spectral weight to form a new parabolic-like band. Each step of this argument depends only on the existence of a Mott / CT gap and on the locality of the dipole operator, neither of which is specific to dimensionality or to the precise microscopic dressing of the equilibrium Hubbard manifolds (the holon–spinon kinematic continuum in 1D, the AFM spin polaron in 2D), so the picture carries over from 1D to 2D unchanged.

\begin{figure*}[htb]
	\centering
	\includegraphics[width=\textwidth]{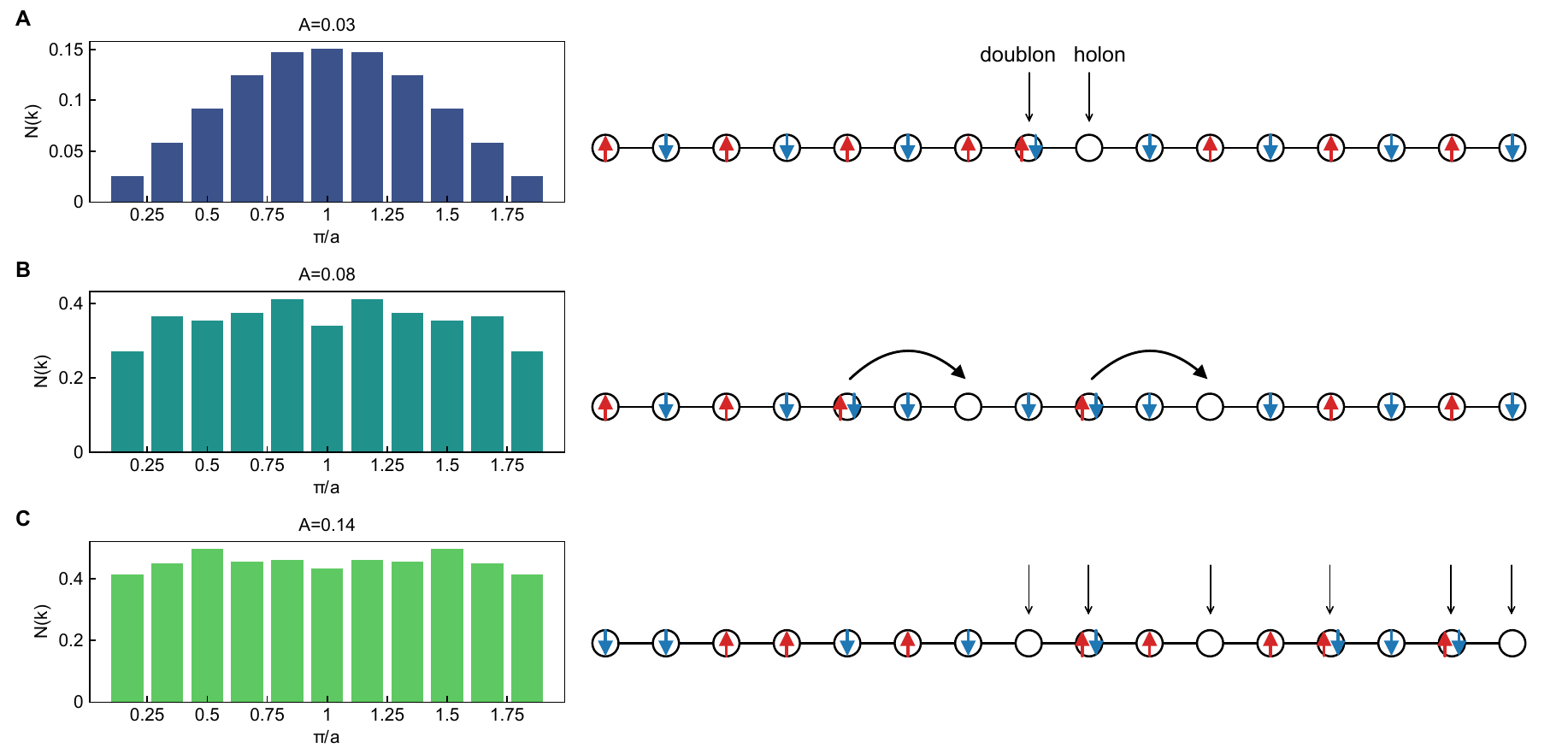}
	\caption{\textbf{Instantaneous charge structure factor after the pump.} Left: charge structure factor $N(\mathbf{k})$ as a function of crystal momentum $k$ (in units of $\pi/a$) for three representative pump amplitudes (\textbf{A}) $A=0.03$, (\textbf{B}) $A=0.08$, and (\textbf{C}) $A=0.14$. Right: schematic real-space configurations of the corresponding many-body states, with arrows denoting singly occupied sites (red: spin up; blue: spin down), doubly occupied sites (doublons) and empty sites (holons) shown as paired ↑↓ and empty circles, respectively. (\textbf{A}) In the weak-drive regime, the ground state remains close to half-filling with a single bound doublon–holon pair, and $N(\mathbf{k})$ is peaked near $k=\pi/a$. (\textbf{B}) At intermediate drive, additional doublon–holon pairs are created; curved arrows indicate the field-driven hopping that separates each doublon from its partner holon, leading to a broader, flatter $N(\mathbf{k})$ whose peak shifts away from $\pi/a$. (\textbf{C}) For strong drive, multiple doublons and holons are generated and propagate throughout the chain (vertical arrows mark their positions), resulting in a nearly featureless $N(\mathbf{k})$ characteristic of a highly excited, spin-disordered state.
    }
	\label{sfig:hubbard charge structure factor}
\end{figure*}

To verify this picture in our simulation, we computed the instantaneous charge structure factor
\begin{equation}
    N(\mathbf{k},t)=\langle\psi(t)|\rho_\mathbf{-k}\rho_\mathbf{k}|\psi(t)\rangle
\end{equation}
and $\rho_\mathbf{k}=\sum_{\mathbf{q},\sigma}c^\dagger_{\mathbf{k+q},\sigma}c_{\mathbf{q},\sigma}$ is the charge density operator. This quantity measures how strongly charge-density fluctuations are correlated at a momentum $\mathbf{k}$, revealing charge ordering tendencies and spatial patterns of density fluctuations. Specifically, $N(\pi)=\frac{1}{L}\langle(\sum_j(-1)^jn_j)^2\rangle$ measures fluctuations of the staggered charge imbalance and grows large when there are strong local staggered doublon-holon correlations. Figure~\ref{sfig:hubbard charge structure factor} shows the momentum distribution of $N_\mathbf{k}$ after the pump pulse for three representative fluences. At low fluence ($A_0=0.03$), $N_\mathbf{k}$ is peaked at $k=\pi$, consistent with the local, nearest-neighbor pair structure of the photo-generated doublon-holon population (Fig.~\ref{sfig:hubbard charge structure factor}A). At the critical fluence ($A_0=0.08$), the peak of $N_\mathbf{k}$ broadens and shifts towards the neighboring momenta of $k=\pi$, suggesting that doublons and holons begin to separate from each other after creation (Fig.~\ref{sfig:hubbard charge structure factor}B). Going far beyond the threshold ($A_0=0.14$), $N_\mathbf{k}$ is overall flat with peaks and humps away from $k=\pi$, reflecting a configuration of randomly distributed doublons and holons (Fig.~\ref{sfig:hubbard charge structure factor}C). These observations directly support the universal picture above: the spatial pair correlation that characterizes the low-fluence regime collectively dissolves across $A_{0c}$.

The sharpness of the crossover at $A_{0c}$, however, does depend on dimension, and we trace this to a simple geometric counting of pair-breaking pathways. Consider a doublon adjacent to a holon on the otherwise half-filled background. The doublon has $z = 2$ nearest neighbors on the 1D chain and $z = 4$ on the 2D square lattice. One of these is the holon, and a hop onto it would annihilate the pair; the remaining $z - 1$ sites are singly occupied and available for intraband hopping. A doublon paired with its holon partner therefore has $z - 1 = 1$ intraband channel in 1D and $z - 1 = 3$ in 2D, while once the doublon has moved away from its partner the number of available channels grows to $z = 2$ in 1D and $z = 4$ in 2D. Therefore, the contrast between the paired and separated configurations is larger in 1D, and the band renormalization effects across $A_{0c}$ are sharp. In 2D, the doublon already has substantial intraband freedom while paired, and the transition is correspondingly smoother and broadened over a continuous range of fluences, though a threshold remains visible. This expectation is consistent with our observations: the 1D simulation (Fig.~\ref{fig:m4}F–H, Fig.~\ref{sfig:hubbard dos}) exhibits a sharp downward evolution of the parabolic-like UHB feature across $A_{0c}$, whereas the 2D experiment (Fig.~\ref{fig:m3}B) shows a smoother onset of below-gap HHG saturation or suppression as fluence grows. The qualitative mechanism — independent propagation of doublons and holons above $A_{0c}$, manifesting as a parabolic-like feature in the UHB spectral function — applies to both 1D and 2D, while the observed difference in crossover sharpness is qualitatively consistent with the geometric counting above.

\clearpage

\end{document}